\documentclass[12pt]{iopart}
\pdfoutput=1
\usepackage{amssymb}
\usepackage{epsfig}
\usepackage{subfigure}
\usepackage{graphicx,color}
\usepackage{textcomp}
\usepackage{cite}
\usepackage{float}
\usepackage{url}
\usepackage{amsopn}
\DeclareMathOperator{\erf}{erf}
\DeclareMathOperator{\erfc}{erfc}


\expandafter\let\csname equation*\endcsname\relax
\expandafter\let\csname endequation*\endcsname\relax
\usepackage{amsmath}

\begin{document}
\title{Stochastic Search with Poisson and Deterministic Resetting}

\author{Uttam Bhat} 
\address{Department of Physics, Boston University, Boston, MA 02215, USA}
\address{Santa Fe Institute, 1399 Hyde Park Road, Santa Fe, NM 87501, USA}
\author{Caterina De Bacco} 
\address{Santa Fe Institute, 1399 Hyde Park Road, Santa Fe, NM 87501, USA}
\author{S. Redner}
\address{Santa Fe Institute, 1399 Hyde Park Road, Santa Fe, NM 87501, USA}

\begin{abstract}

  We investigate a stochastic search process in one, two, and three
  dimensions in which $N$ diffusing searchers that all start at $x_0$ seek a
  target at the origin.  Each of the searchers is also reset to its starting
  point, either with rate $r$, or deterministically, with a reset time $T$.
  In one dimension and for a small number of searchers, the search time and
  the search cost are minimized at a non-zero optimal reset rate (or time),
  while for sufficiently large $N$, resetting always hinders the search.  In
  general, a single searcher leads to the minimum search cost in one, two,
  and three dimensions.  When the resetting is deterministic, several
  unexpected feature arise for $N$ searchers, including the search time being
  independent of $T$ for $1/T\to 0$ and the search cost being independent of
  $N$ over a suitable range of $N$.  Moreover, deterministic resetting
  typically leads to a lower search cost than in stochastic resetting.

\end{abstract}


\section{Introduction}

Stochastic searching~\cite{BLMV11} underlies many biological
processes~\cite{BWV81,V07,M08}, animal
foraging~\cite{C76,B91,OBE90,VBH99,BLMV06}, as well as operations to find
missing persons or lost items~\cite{RS71,FS01,S09}.  In these settings basic
goals are to maximize the probability that the target is actually found and
to minimize the time and/or the cost required to find the target.  In
response to these challenges, a wide variety of search algorithms have been
extensively investigated and rich dynamical behaviors have been
uncovered~\cite{MOS11,MMMO12}.

Typically, one or perhaps multiple searchers move in some fashion through a
search domain to locate either a single target or a series of targets.  The
most naive setting is that of a single searcher has no information about the
target and moves by random-walk, or equivalently, diffusive motion.  Such a
search is generally hopelessly inefficient because the target may not be
found, for spatial dimension $d\geq 3$, or the average search time is
infinite.  Thus much effort has been directed to uncover more effective
search strategies

Many such possibilities have been investigated.  One natural mechanism is to
allow the searcher to move according to a L\'evy flight (see, e.g.,
\cite{VBH99,SK86,VLRS11}), so that the search can quickly cover large
distances between targets.  A somewhat related example is that of
intermittent search, in which the search process is partitioned into periods
of intensive search, during which the searcher moves slowly, and superficial
search, during which the searcher moves quickly~\cite{BCMSV05}.  In the
context of searching for nourishment, the essential tradeoff is how long to
continue to exploit resources in a local area and when to move to a new area
as local resources become depleted~\cite{C76,BR14,CBR16}.  These notions also
underlie the search for a target along a DNA by a diffusing
protein~\cite{BWV81,V07,WV81,WBV81,HM04,HGS06}, where the tradeoff is for the
search to diffuse along the DNA or unbind and reattach at some distant point
along the DNA.

Very recently, the mechanism of search that is augmented by ``resetting'' was
introduced~\cite{EM11a,EM11b,EM14}.  In this model, a target is placed at the
origin (without loss of generality) and a searcher starts at some arbitrary
point.  In addition, the searcher returns to a fixed ``home base'' at a given
rate during the search.  If the distance between the home base and the target
is known with certainty, then a search based on a stochastically moving
searcher is not a pertinent approach.  However, the natural situation is that
the target location is only partially known; for example, the target is
somewhere within a finite body of water.  In this case, a relevant parameter
is the maximum possible distance between the target and the home base.  As
shown in~\cite{EM14}, the basic properties of search with resetting when the
distance between the target and home base is known precisely are
qualitatively the same as the situation where only the probability
distribution of this distance is known.  Thus for the purposes of
tractability we restrict ourselves to the idealized (and admittedly
unrealistic) situation where the distance between the target and home base is
known.

In general, resetting is known to have a dramatic effect on the search.  A
diffusing particle requires an infinite average time to reach a target in
spatial dimensions $d=1$ and $d=2$, and the searcher may not even reach a
finite-size target for $d>2$.  However, resetting ensures that: (i) the
searcher can always find the target in any dimension and (ii) the average
search time is finite.  Overall, therefore, resetting gives rise to a more
efficient search.  One of the basic results of recent investigations of
search with resetting~\cite{EM11a,EM11b,EM14} was to determine the conditions
that optimize the search time.  This resetting mechanism has also been quite
fruitful conceptually and a variety of interesting consequences of resetting
have been
elucidated~\cite{MV13,AG13,BS14,CS15,CM15,MSS15,P15,KG15,TRU15,EM16}.

\begin{figure}[ht]
\centerline{\includegraphics[width=0.6\textwidth]{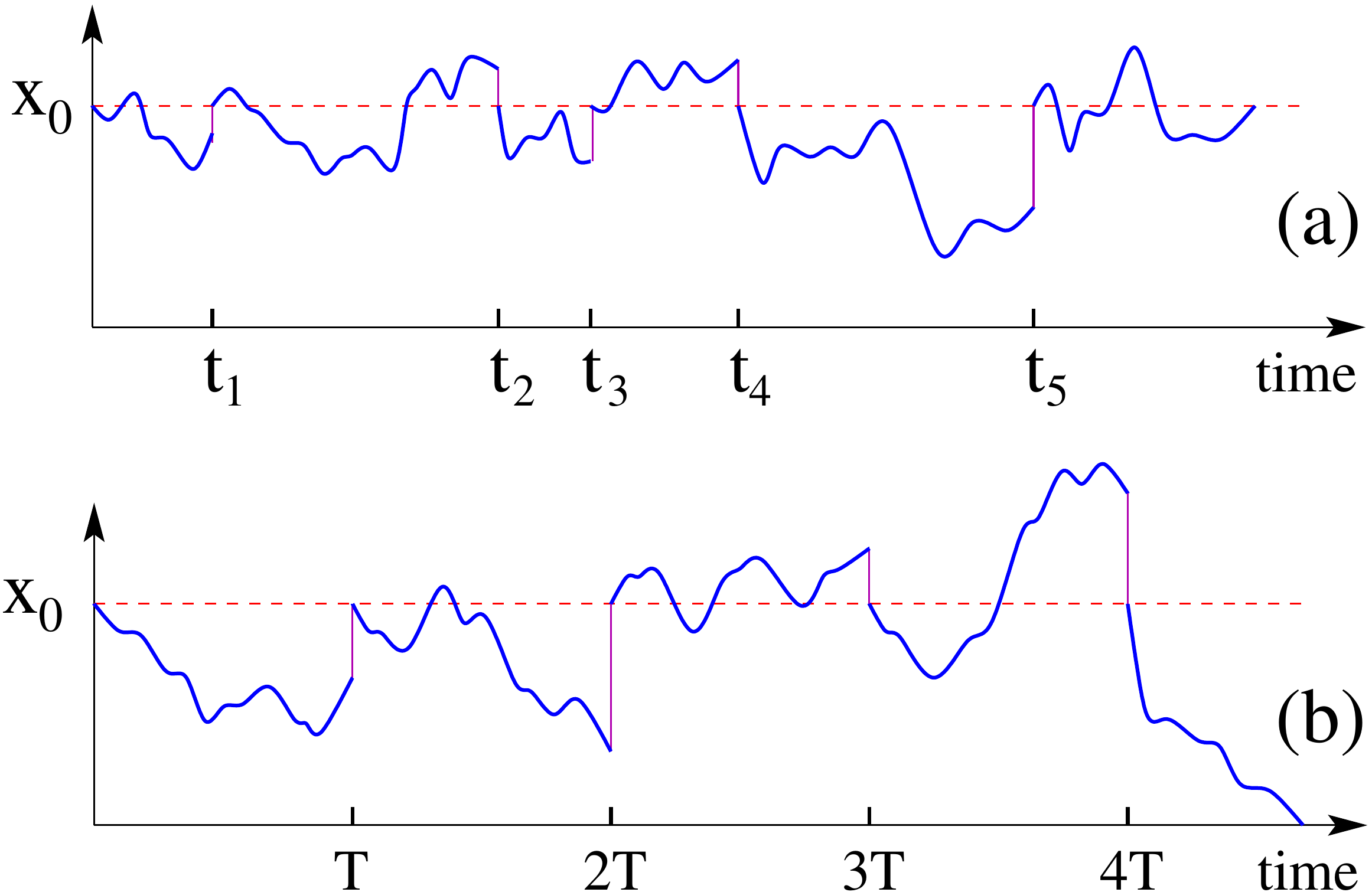}}
\caption{Contrast of the trajectories in: (a) Poisson reset and (b) deterministic
  reset with reset time $T$.  The target is at the origin and the reset point
  is $x_0$.}
\label{1d-illust}
\end{figure}

In this work, we investigate two as yet unexplored features of search with
resetting: (i) $N$ searchers, each of which is reset at the same Poisson
rate, to a ``home base'', and (ii) deterministic reset, in which the
searchers return to the home base after a fixed operation time, rather than
the searchers being reset according to a fixed-rate Poisson process
(Fig.~\ref{1d-illust}).  The related situation of many searchers that are
uniformly distributed in space, each of which is reset to its own starting
position at a fixed rate, was investigated in~\cite{EM14}.  However, in the
context of search for a missing person, it is natural that all the searchers
return to a single home base (or perhaps a small number of such bases).
Moreover, in such a search, activities are typically suspended at the end of
daylight or when searchers reach their physical limits.  Thus it is also
realistic to investigate the situation in which all the searchers are reset
to a given location at a fixed reset time.

In the next section, we start by briefly reviewing known results about
stochastic search by a single searcher in one dimension ($d=1$), with the
additional feature that the searcher is reset to its home base at a fixed
rate.  We will also present our renewal-based approach to solve this problem
that will be employed throughout this work.  Next, we treat the case of $N$
searchers in one dimension, each of which is independently reset to the same
location at a fixed rate $r$.  We show that for $N<N^*$ there is an optimal
non-zero reset rate $r^*_N$ that minimizes the search time as well as the
search cost, while for $N>N^*$ resetting always hinders the search.  We also
determine this critical number $N^*$ analytically.  In Sec.~\ref{sec:d>1}, we
turn to search with stochastic resetting in spatial dimensions $d=2$ and
$d=3$.  For the case of $d=3$, we exploit a well-known construction to reduce
the diffusion equation in three dimensions to an effective diffusion equation
in one dimension.  This allows us to obtain results about three-dimensional
search in terms of the corresponding one-dimensional system.

To probe the properties of the $N$-searcher system in a convincing way, we
outline, in Sec.~\ref{sec:EDS}, an efficient event-driven simulation in one
dimension that obviates the need to microscopically follow the trajectories
of each searcher between reset events.  By exploiting the aforementioned
dimensional reduction of the three-dimensional diffusion equation, we can
also directly adapt our event-driven approach to three dimensions.  For
$d=2$, no such dimensional reduction exists and our simulations are based on
a more direct approach.  From these numerical approaches, we determine the
condition for optimal search for both a single searcher and for many
searchers in spatial dimensions $d=1$ and 3, and then $d=2$.

Finally, in Sec.~\ref{sec:DR}, we investigate the case of deterministic
resetting, for both a single and for many searchers, again for the cases of
$d=1$, $d=3$, and then $d=2$.  The salient feature of deterministic resetting
is that it leads to a quicker search than stochastic resetting at their
respective optimal resetting rates (or times).  Moreover, deterministic
resetting leads to a search cost that, for large $N$, is nearly independent
of the reset rate $r$ over a wide range of $r$.  Concluding remarks are given
in Sec.~\ref{sec:disc}.

\section{Poisson Resetting in One Dimension}
\label{sec:SR1d}

As mentioned above, the situation where a searcher is reset at a fixed rate
$r$ has already been extensively investigated~\cite{EM11a,EM11b,EM14}.  For
completeness, we quote the main results for this type of search and also
derive them by an independent method.  We then investigate the case of $N$
independent searchers, each of which is reset to a common point at the same
rate $r$.

\subsection{One searcher}

Consider a target that is fixed at the origin and a diffusing searcher that
is reset to a point $x_0$ at rate $r$.  For simplicity, we assume that the
searcher begins at this reset point.  For this system, the first two moments
of the search time are (see Refs.~\cite{EM11a,EM11b} and also \ref{PD})
\begin{align}
\begin{split}
\label{t}
\langle t_1\rangle &=\frac{1}{r}\left(e^{x_0 \sqrt{{r}/{D}}}-1\right)\,,\\
\langle t_1^2\rangle &= \frac{1}{r^2}\left[e^{x_0 \sqrt{{r}/{D}}} \big(2\, e^{x_0
   \sqrt{{r}/{D}}}-x_0 \sqrt{{r}/{D}}-2\big)\right]\,,
\end{split}
\end{align}
and higher moments can be extracted straightforwardly.  The subscript 1
signifies one searcher.  The basic feature of \eqref{t} is that
$\langle t_1\rangle$ is minimized at an optimal reset rate $r^*$ that is of
the order of $D/x_0^2$.  The inverse of the optimal rate gives the typical
time between resets as roughly $T_D\equiv x_0^2/D$, the time for a diffusing
particle to reach a distance $x_0$.  If the searcher does not find the target
within this time, then it is likely wandering in the wrong direction and will
reach the target at a time much greater than $T_D$.  In this case, it is
better to reset this errant searcher back to its home base than allowing it
to continue on its current trajectory.

We now give an independent derivation for the average search time
$\langle t_1\rangle$ that relies on the renewal nature of the search process;
a similar approach was very recently developed in Ref.~\cite{R16}.  Namely,
whenever a reset occurs, the process restarts at the initial condition, but
with the proviso that the time is incremented appropriately to account for
the return to the reset point.  As a preliminary, we need the following:
\begin{align}
R(t)&\equiv\text{prob.\ reset time is greater than}\ t\,,\nonumber\\
S(x_0,t)&\equiv\text{prob.\ hitting time is greater than}\ t\,.\nonumber
\end{align}
For diffusive motion, $S(x_0,t)$ is also the ``survival probability'' that a
searcher initially at $x_0$ has not reached the origin by time $t$
is~\cite{R01}
\begin{subequations}
\begin{equation}
\label{S}
S(x_0,t)= \text{erf}\Big(\frac{x_0}{\sqrt{4 D t}}\Big)\,.
\end{equation}
Correspondingly, the first-passage probability that a diffusing particle
initially at $x_0$ first reaches the origin at time $t$ is
\begin{equation}
\label{FPP}
F(x_0,t)=-\frac{d S(x_0,t)}{dt}= \frac{x_0}{\sqrt{4\pi D t^3}}\,\, e^{-x_0^2/4Dt}\,\,.
\end{equation}
\end{subequations}

Since the search process is specified by whether the target is reached before
a reset occurs and vice versa, we also define  two fundamental probabilities: 
\begin{align}
\begin{split}
\label{q}
P&\equiv\text{prob.\ target hit before reset}
 =\int_0^\infty\!\!dt
\underbrace{~\left(-\frac{dS(x_0,t)}{dt}\right)~}_{\text{prob.\ hit 
    in $(t,t+dt)$}}\times\
\underbrace{~~~R(t)~~~}_{\text{prob.\
      reset time $>t$}}\!\!\!\!\!,\\
Q&\equiv\text{prob.\ reset before target hit}
=\int_0^\infty\!\!dt  \underbrace{~~S(x_0,t)~~}_{\text{prob.\ hit time $>t$}}\,\times\, 
\underbrace{~~\left(-\frac{dR(t)}{dt}\right)~~}_{\text{prob.\
      reset in $(t,t\!+\!dt)$}}\!\!\!\!.
\end{split}
\end{align}
Thus the ``direct'' time for the target to be reached before a reset occurs,
which we define as $t_d$, is
\begin{subequations}
\begin{align}
\label{tdtr}
t_d  =\int_0^\infty\!\!dt\, t\,
\left(-\frac{dS(x_0,t)}{dt}\right)\,\times\, R(t)\Big/P\,.
\end{align}
Similarly, the ``reset'' time $t_r$ for a reset to occur before the target is
reached is
\begin{align}
t_r  =\int_0^\infty\!\!dt\, t\,
\left(-\frac{dR(t)}{dt}\right)\,\times\, S(x_0,t)\Big/Q\,.
\end{align}
\end{subequations}

Using the renewal nature of the search, $\langle t_1\rangle$ satisfies the
recursion
\begin{subequations}
\begin{equation}
\langle t_1\rangle= Pt_d+Q\big(t_r+ \langle t_1\rangle\big)\,. 
\end{equation}
The first term accounts for hitting the target before a reset occurs, while
the second term accounts for the search restarting after a reset.  In this
latter case, the search is delayed by $T_r$.  For notational convenience, we
define $T_d=P\,t_d$ and $T_r=Q\, t_r$.  Solving for $\langle t_1\rangle$
gives
\begin{equation}
\label{t-recur}
  \langle t_1\rangle= \frac{T_d+T_r}{1-Q}\,\,.
\end{equation}
\end{subequations}
This result is general and can be applied to higher dimensions and to
different reset mechanisms, as will be discussed later.

Since only the sum $T_d+T_r$ appears in the expression for
$\langle t_1\rangle$, we add the two lines in \eqref{tdtr} and integrate by
parts to give
\begin{align}
\label{TdTr}
T_d+T_r= \int_0^\infty dt\, R(t)\, S(x_0,t)\,.
\end{align}

We now specialize to Poisson resetting with rate $r$, for which
$R(t)=e^{-rt}$.  Using this, as well as \eqref{S} for $S(x_0,t)$, the
integrals in \eqref{q} and \eqref{TdTr} are
\begin{align}
T_d+T_r&= \frac{1}{r}\!\left(1\!-\!e^{-x_0\sqrt{r/D}}\right)\,, \nonumber\\
Q&=  1-e^{-x_0\sqrt{r/D}}\,.\nonumber
\end{align} 
from which \eqref{t-recur} gives
\begin{align}
\label{t-recur-b}
\langle t_1\rangle =\frac{1}{r}\left(e^{x_0\,\sqrt{r/D}}-1\right)\,.
\end{align}
This reproduces Eq.~\eqref{t} as it must.

\subsection{Multiple searchers}

Since all searchers are independent and the target location is fixed, it is
theoretically possible to obtain the survival probability of the target in
the presence of $N$ searchers as the $N^{\rm th}$ power of the target
survival probability due to a single searcher.  However, while the Laplace
transform of the target survival probability with one searcher is known
exactly, Eq.~\eqref{S1s}, it does not appear possible to Laplace invert this
expression exactly.  Nevertheless, we can invert this Laplace transform in
the limit $r\to 0$ to provide information about the dependence of the search
cost as $r\to 0$; this feature will be discussed below.

Thus we resort to simulations to map out the behavior of the search time as a
function of the reset rate $r$ for multiple searchers.  Because it is
inherently wasteful to simulate directly the microscopic motion of each
searcher between reset events, we developed an efficient event-driven
simulation, whose details are given in Sec.~\ref{sec:EDS}.  Our focus is on
the rich features of the search time and the search cost as a function of $r$
and $N$.  Under the assumption that each searcher has the same fixed cost per
unit time of operation, the search cost for $N$ searchers, $C_N$, is merely
$C_N=N\langle t_N\rangle$, where $\langle t_N\rangle$ is the average search
time for $N$ searchers.  This cost has a weak dependence on $N$, so that it
is more convenient to focus on cost rather than time in the following.

\begin{figure}[ht]
\centerline{\subfigure[]{\includegraphics[width=0.5\textwidth]{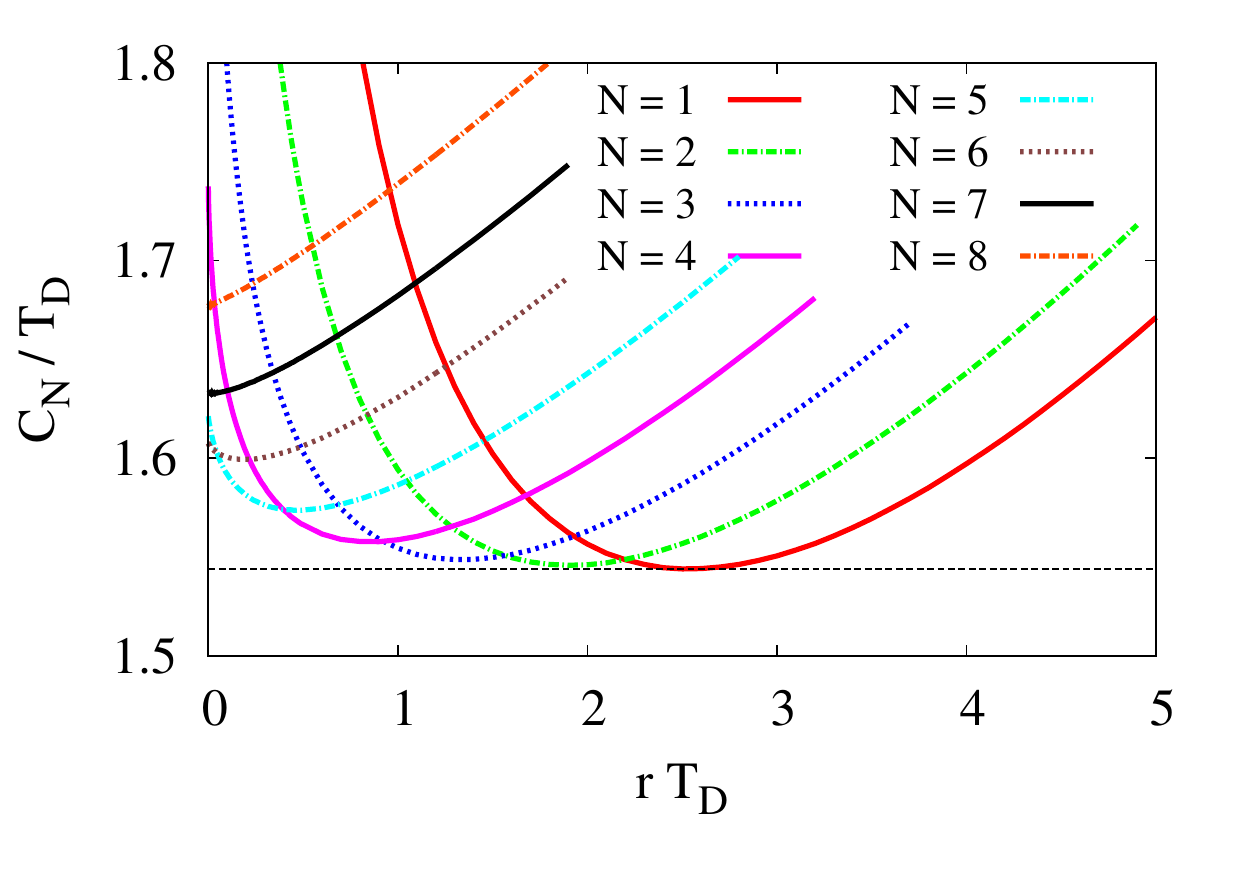}}
\subfigure[]{\includegraphics[width=0.48\textwidth]{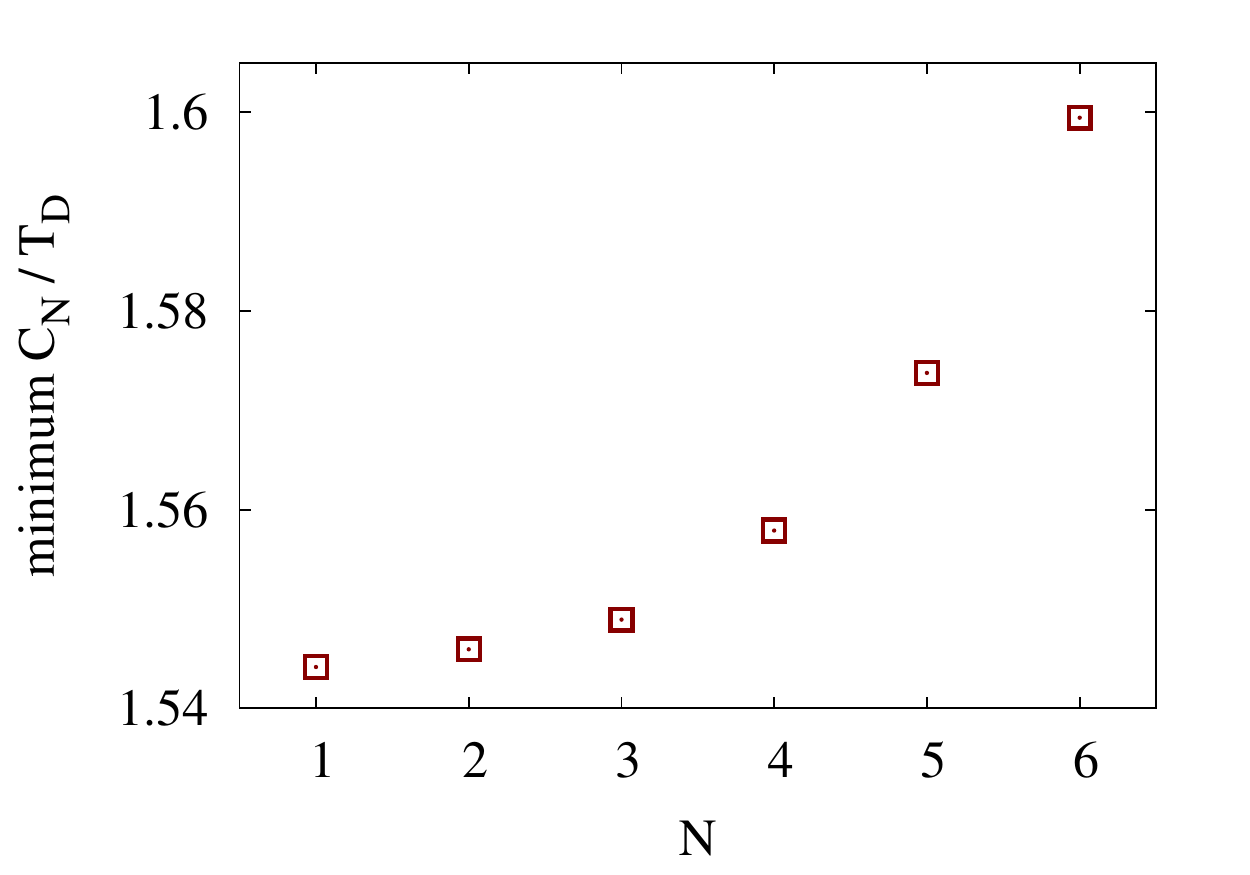}}}
\caption{(a) Average search cost $C_N$ scaled by the diffusion time $T_D$
  versus the scaled reset rate $rT_D$ in one dimension for various $N$.  Each
  curve represents an average over $10^9$ trajectories.  The dashed line
  indicates the minimum cost of 1.544 for $N=1$.  (b) The minimum search cost
  for each $N\leq 6$.}
\label{1Dsim}
\end{figure}

Figure~\ref{1Dsim} shows the search cost in one dimension as a function of
$r$ for $1 \leq N\leq 8$.  Noteworthy features of the search cost include:
\begin{enumerate}

\item The lowest scaled search cost of $1.544$ is achieved by a single
  searcher that is reset at the scaled optimal rate $r^*_1 \approx 2.540$.
  At this optimum, there are typically $r^*_1\langle t_1\rangle\approx 3.657$
  resets before the searcher reaches the target.

\item For $N=2,3,\ldots 7$, there is a unique, non-zero optimal reset rate
  $r^*_N$ for each $N$ that minimizes the search cost and the search time.
  This optimal rate is generally of the order of the inverse diffusion time
  between the home base and the target, $T_D=x_0^2/D$.  The optimal cost for
  all $N\leq 5$ is within 2\% of the optimal cost for a single searcher.

\item For $N\!\geq\! 8$, the search time strictly increases with $r$; that
  is, no resetting is optimal.  This behavior arises because at least one
  searcher is systematically moving toward the target, once the number of
  searchers is sufficiently large, so that any resetting increases the search
  time.  The demonstration of the sign change in the initial slope of
  $\langle t_N\rangle$ versus $r$ between $N=7$ and $8$ (see
  Fig.~\ref{fig:dtdr}) is given in \ref{7-8}.

\begin{figure}[ht]
\centerline{\includegraphics[width=0.5\textwidth]{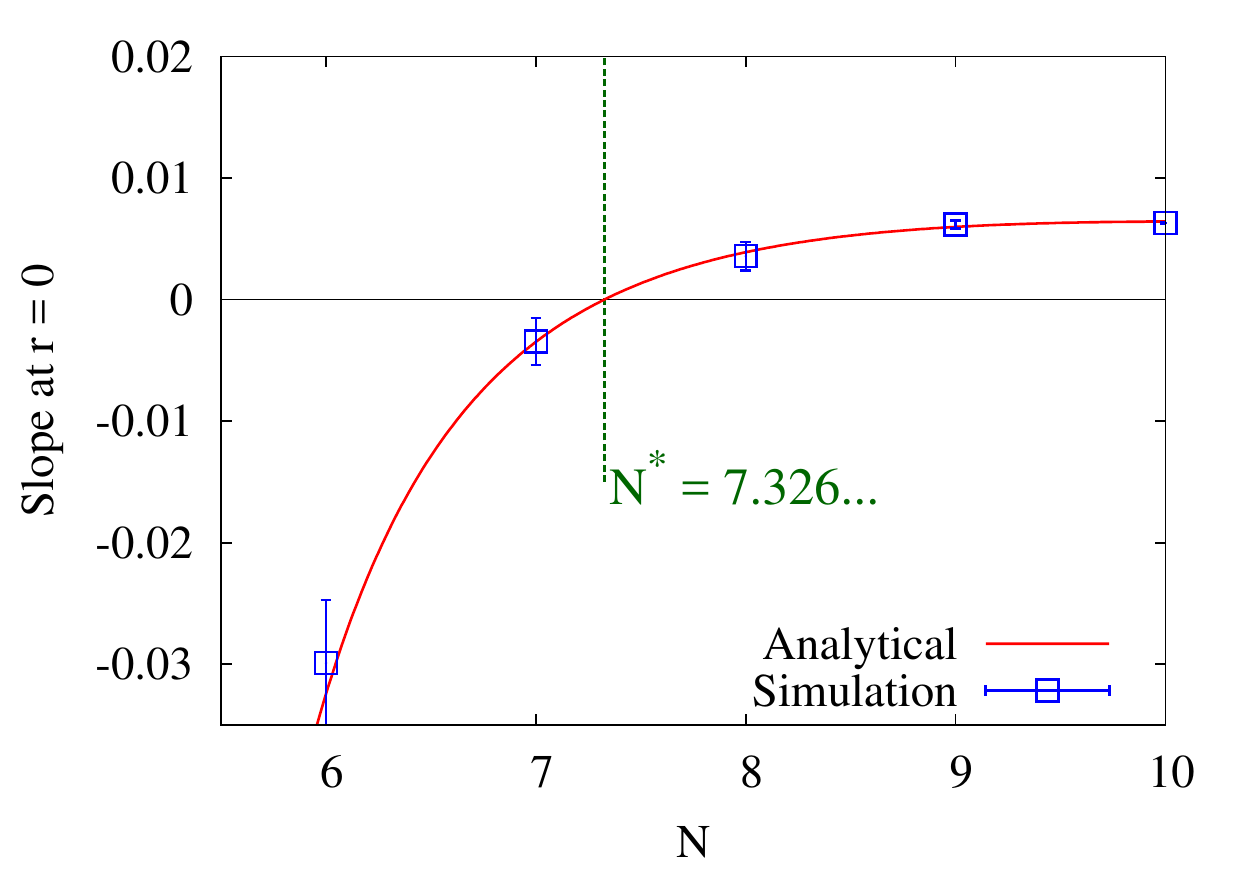}}
\caption{ Slope of the average search time $\langle t_N\rangle$ at $r=0$ as a
  function of $N$ in one dimension when each searcher is independently reset
  at rate $r$.}
\label{fig:dtdr}
\end{figure}

\item For $N=1$, the average search time diverges as $r\to 0$.  This property
  reflects the divergence of the first-passage time for a diffusing particle
  to hit an arbitrary point in one dimension~\cite{F68,R01}.  Since the
  survival probability for the diffusing particle to not hit the target by
  time $t$, $S(x_0,t)$ in Eq.~\eqref{S}, asymptotically decays as $t^{-1/2}$,
  we estimate the hitting time for small $r$ as
  $\int^{1/r} S(x_0,t)\, dt\sim r^{-1/2}$.  This reproduces the behavior that
  arises from a small-$r$ expansion of search time in Eq.~\eqref{t}.

\item For $N=2$, the search time again diverges as $r\to 0$.  Because of the
  independence of the searchers, the survival probability of the target
  asymptotically decays as $\big[t^{-1/2}\big]^2=t^{-1}$.  In the $r\to 0$
  limit, the same argument as that given for $N=1$ leads to an average search
  time that diverges as $-\ln r$ for $r\to 0$.

\item The probability that the target is not hit by any of $N$ searchers
  asymptotically decays as $\big[t^{-1/2}\big]^N$.  Thus when there are at
  least 3 searchers, the search time is finite for $r\to 0$,

\end{enumerate}

\section{Poisson Resetting in Higher Dimensions}
\label{sec:d>1}

\subsection{Three dimensions}

There are two important physical differences between the one-dimensional and
three-dimensional system: (a) First, the target must have a non-zero size to
be detected; we take the target to be an absorbing sphere of radius $a$.  (b)
Second, the existence of a non-zero target radius introduces an additional
parameter---the ratio of the target radius to the radius of the reset point
$a/r_0$.

In spite of these two complications, the above approach for one dimension can
be straightforwardly adapted to three dimensions because of the well-known
correspondence between the diffusion equation in three dimensions and in one
dimension~\cite{S43,R01}.  Namely, the three-dimensional radial Laplacian
operator is related to the one-dimensional Laplacian by
$r\,\nabla^2_{3d}\, P= \nabla^2_{1d}\,(rP)$.  Using this mapping, we can
write basic quantities for search with resetting in three dimensions in terms
of corresponding one-dimensional expressions.

\begin{figure}[ht]
\centerline{\subfigure[]{\includegraphics[width=0.50\textwidth]{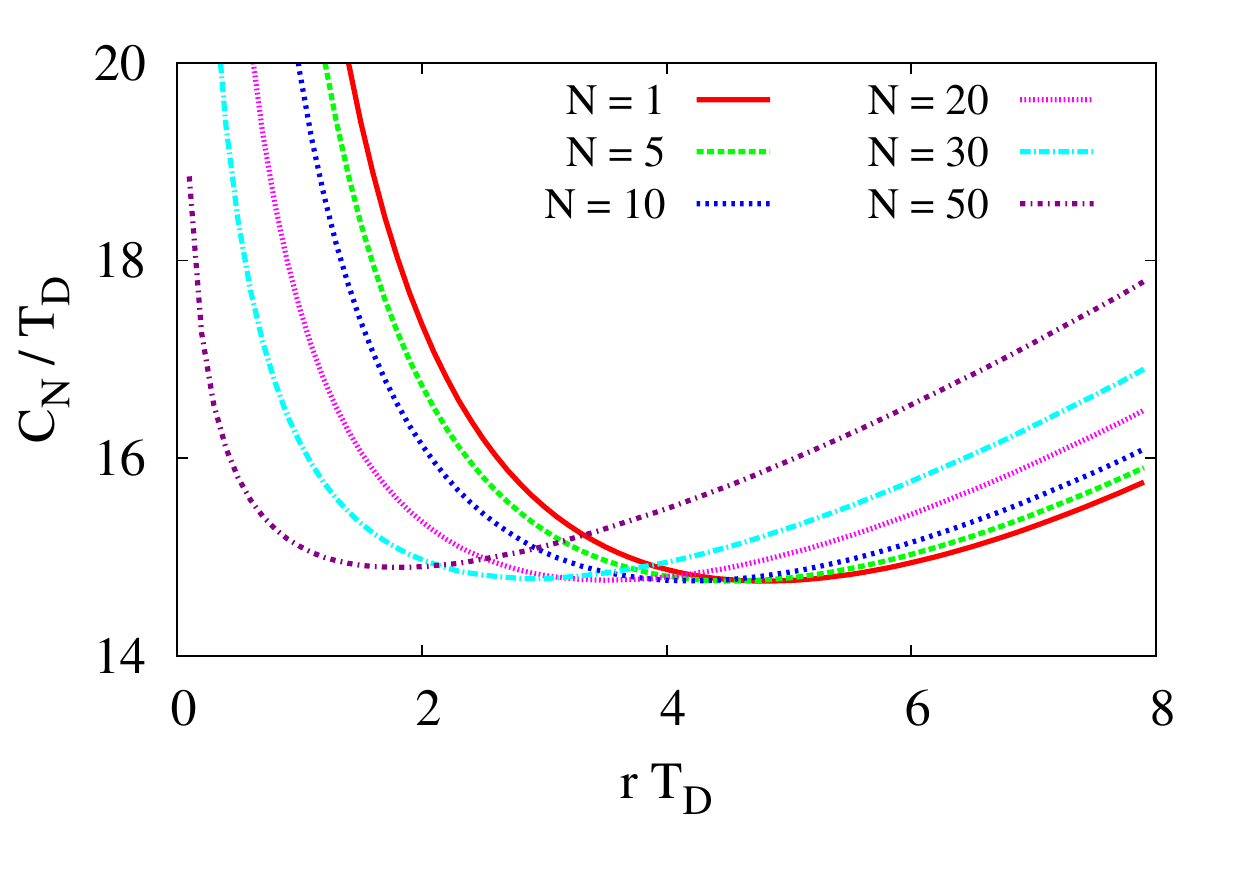}}
\subfigure[]{\includegraphics[width=0.48\textwidth]{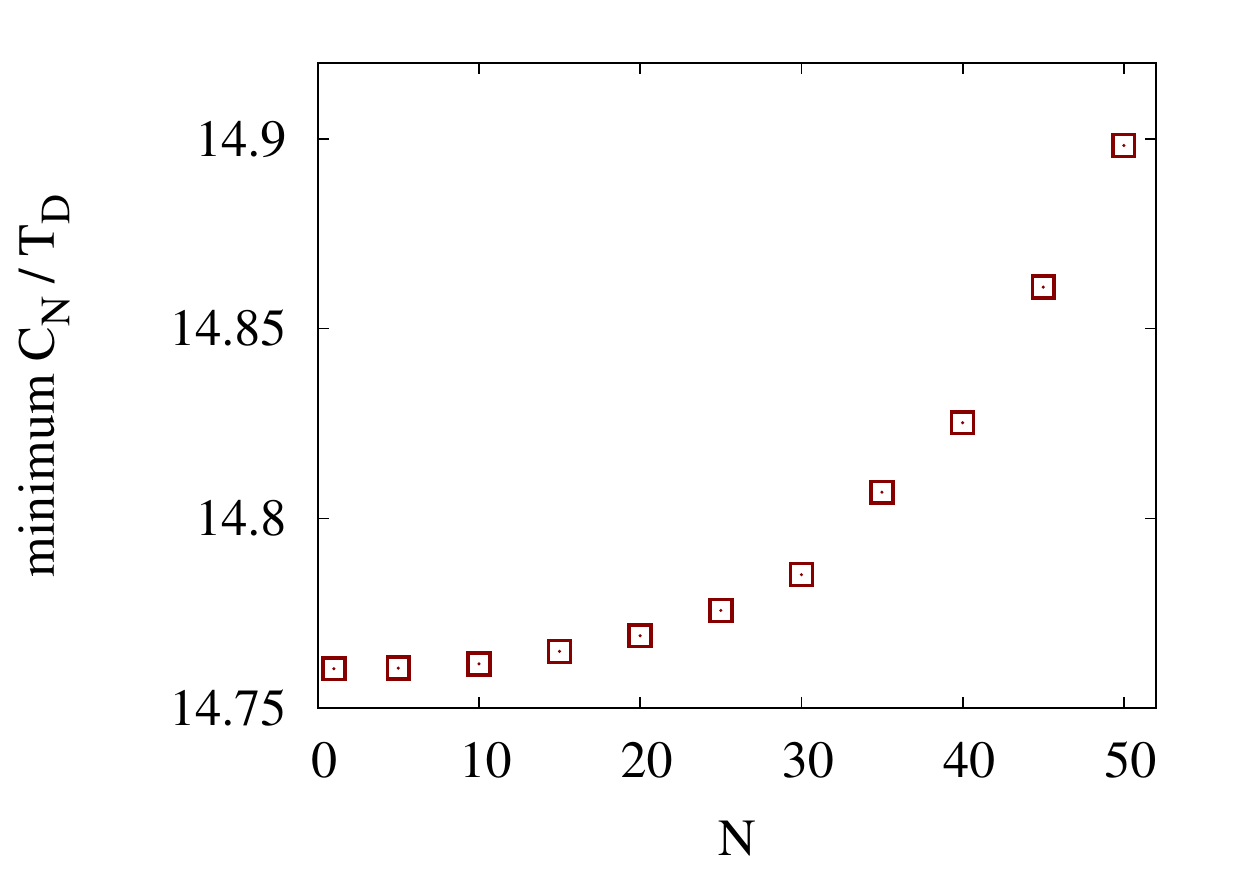}}}
\caption{(a) Average scaled search cost $C_N/T_D$ versus scaled reset rate
  $rT_D$ in three dimensions for various $N$ and for $a/r_0=10^{-1}$.  The
  curves for $N = 1$ and 5 are averaged over $10^8$ trajectories, while those
  for $N = 10, 20, 30$ and $50$ are averaged over $10^7$ trajectories.  (b)
  Minimum search cost as a function of $N$.}
\label{3d-stoch}
\end{figure}

Because of computational limitations, most of our numerical results for
stochastic resetting in $d=3$ are for the case of $a/r_0=10^{-1}$
(Fig.~\ref{3d-stoch}).  Simulations for different $a/r_0$ give a
qualitatively similar dependence of the search time and search cost on the
reset rate.  As in the corresponding one-dimensional system, several features
of these results are worth highlighting:

\begin{enumerate}

\item For $N\lesssim 10$, the minimum cost changes so slowly with $N$ that is
  is not possible to determine the value of $N$ at which the minimum cost is
  achieved.  For example, for $N=1$, the minimum cost is $14.760\pm 0.0015$
  while for $N=10$, the minimum cost is $14.762\pm 0.0015$.  By
  $N\approx 15$, however, the minimum cost has a systematic increasing trend
  that is larger than the error bars in the data.

\item In distinction to one dimension, the typical number of reset
  events before the target is found by a single searcher is of the order of
  $r_0/a$, which can be large.  To understand this behavior, we start with
  the hitting probability $H_{\rm 3d}(r_0,t)$ that a single searcher that is
  a distance $r_0$ from the center of the target finds it within time
  $t$~\cite{R01}:
\begin{equation*}
  H_{\rm 3d}(r_0,t)= 1-S_{\rm 3d}(r_0,t)= \frac{a}{r_0}\erfc{\Big(\frac{r_0-a}{\sqrt{4Dt}}\Big)}\,.
\end{equation*}
Here $S_{\rm 3d}(r_0,t)$ is the probability that the searcher does not find
the target within time $t$ (see Sec.~\ref{subsec:3d}).  For $r_0\gg a$ and
reset rate near the optimal value of $D/r_0^2$, the above hitting probability
reduces to $H_{\rm 3d}(r_0,t)\approx \frac{a}{r_0}\text{erfc}(\frac{1}{2})$.
Thus for $a/r_0\ll 1$, the number of reset events until target is found is of
the order of the inverse of this hitting probability, namely, of the order of
${r_0}/{a}$.

\item Because of the transience of diffusion in three dimensions, the search
  time and search cost diverge as $r\to 0$ for any number of searchers.  Thus
  infinitesimal resetting always leads to a more efficient search than no
  resetting.

\end{enumerate}

\subsection{Two dimensions}

\begin{figure}[ht]
\centerline{\subfigure[]{\includegraphics[width=0.50\textwidth]{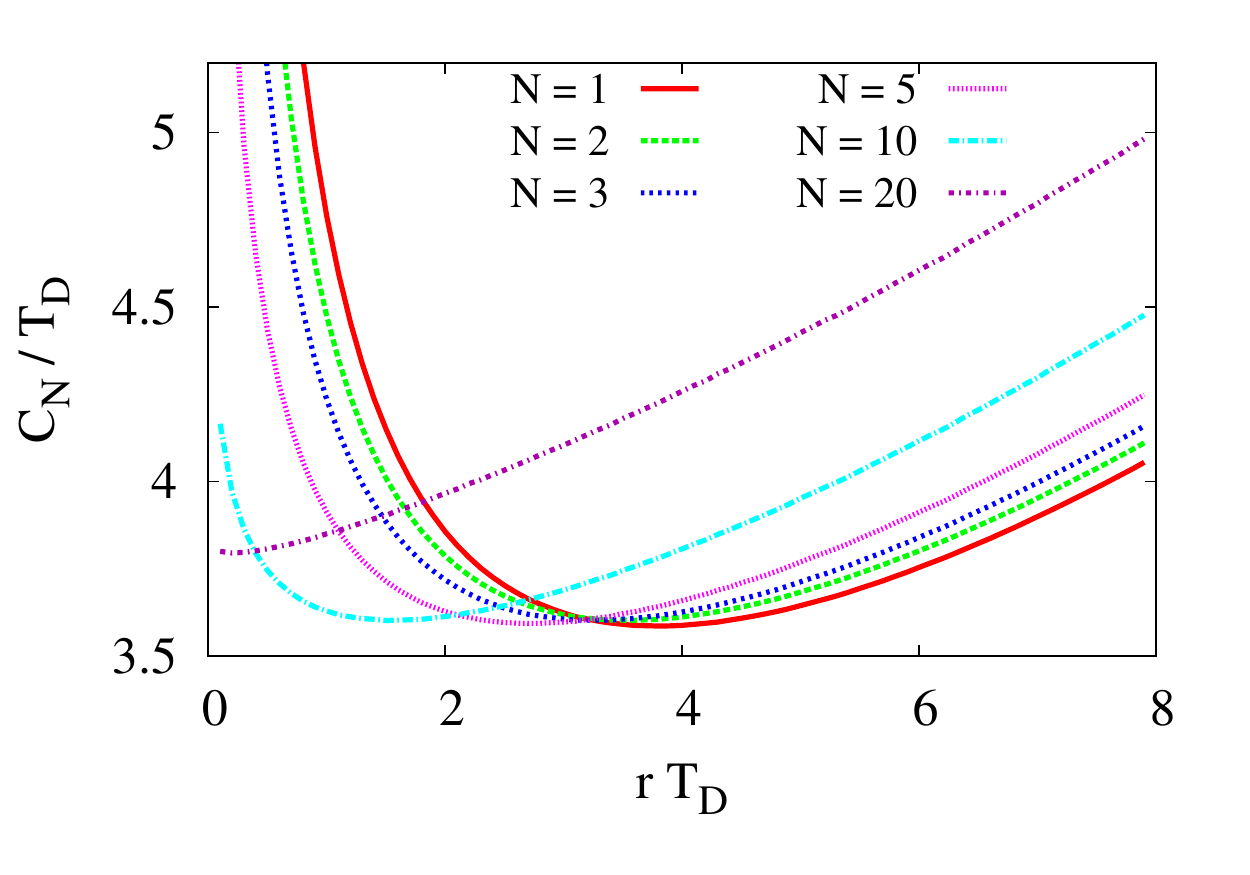}}
\subfigure[]{\includegraphics[width=0.48\textwidth]{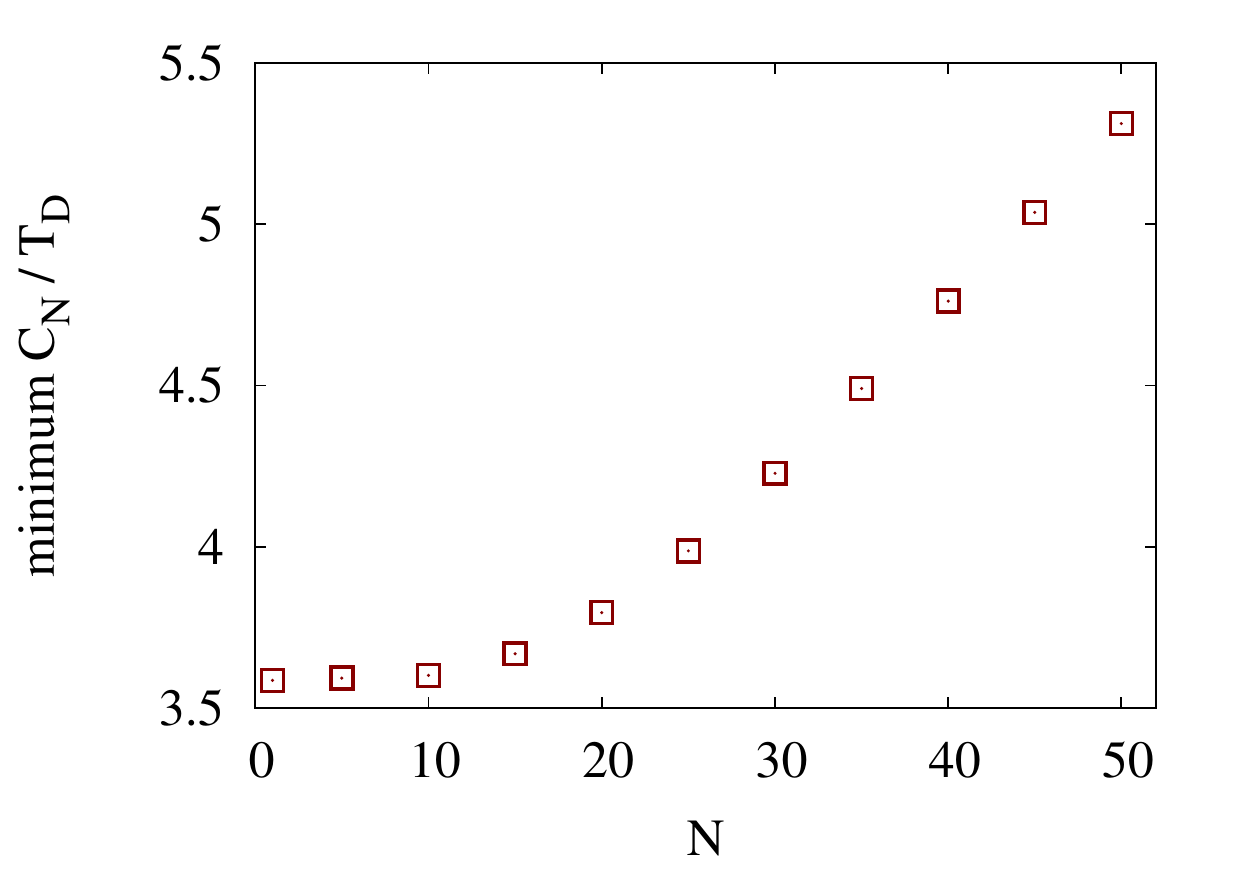}}}
\caption{(a) Average scaled search cost $C_N/T_D$ versus scaled reset rate
  $rT_D$ in two dimensions for various $N$ and for $a/r_0=10^{-1}$.  The data
  are averaged over $10^8$ trajectories.  (b) Minimum search cost as a
  function of $N$. }
\label{2d-stoch}
\end{figure}

In two dimensions, it is practically not feasible to implement an
event-driven simulation because the first-passage and hitting probabilities
that form the kernel of the event-driven algorithm are not known in closed
form as a function of time.  (They are known, however, in the Laplace
domain~\cite{R01}, and were used in~\cite{EM11b,EM14} to provide an
analytical expression for the search time for the case of a single
searcher.)~ Thus we implement an alternative simulational approach, as will
be discussed in the next section.

The primary features of search with stochastic resetting in two dimensions
are (see Fig.~\ref{2d-stoch}):
\begin{enumerate}
\item Resetting again always leads to a more efficient search compared to the
  case of no resetting.  As in three dimensions, this feature is a
  consequence of the divergent average search time at $r=0$ in two dimensions
  for any number of searchers.

\item The dependence of the search cost as a function of reset rate is
  qualitatively similar to that in three dimensions.  The minimum cost is
  nearly constant for $N\lesssim 10$, with the difference in the cost values
  for adjacent $N$ values less than the simulation error bars.  For example,
  the minimum costs for $N=1$ and $10$ are $3.59\pm 0.01$ and $3.60\pm 0.01$
  respectively, while for $N=15$, the minimum cost in $3.67\pm 0.02$.
  
\end{enumerate}

\section{Event-Driven Simulations}
\label{sec:EDS}

The one- and three-dimensional numerical results are based on an event-driven
algorithm that allows us to efficiently simulate $N$ independently resetting
searchers.  In our approach, each searcher is propagated by a single
(typically macroscopic) time step between reset events until one of the
searchers finds the target.  Thus each update is ``useful'' in that either
the target is found or a reset event occurs.  No time is expended in diffusively
propagating searchers between resets.

\subsection{One dimension}

In one spatial dimension, the elemental steps of our algorithm are the
following:
\begin{enumerate}
\item Start with all the searchers at a distance $x_0$ from the target.

\item For each searcher, with the $i^{\rm th}$ one located at $x_i$, draw a
  random time value from the first-passage distribution $F(x_i,t)$ given in
  Eq.~\eqref{FPP}.  Also choose a reset time $t_r$ from a Poisson
  distribution according to the reset rate $Nr$.  This gives the time for one
  of the $N$ searchers to be reset.

\item If the minimum among these $N+1$ times is the reset time, choose a
  random searcher and reset it to $x_0$.  Each of the remaining searchers is
  moved from its current position $x_i$ to a new position that is drawn from
  the conditional probability 
\begin{equation*}
     P(x_i,t_r) = \frac{1}{\sqrt{4 \pi D t_r}} 
\left[e^{-(x_i-x_0)^2/4 D t_r} - e^{-(x_i+x_0)^2/4 D t_r}\right]
\Big/S(x_0,t_r)\,,
\end{equation*}
where $S(x_0,t_r)$ is again the probability that a diffusing particle that
starts at $x_0$ does not hit the target up to time $t_r$.  The distribution
$P(x_i,t_r)$ corresponds to the diffusive propagation of each searcher over
the time increment $t_r$, subject to the constraint that no searcher can
reach the target.  After all the searchers are moved, increment the elapsed
time by the reset time $t_r$ and return to step (i).

\item If the minimum among the $N+1$ random times is one of the first-passage
  times, then the target is found.  The total search time is the current
  elapsed time plus this first-passage time.
\end{enumerate}

\subsection{Three dimensions}
\label{subsec:3d}

By exploiting the dimensional reduction of the diffusion equation in three
dimensions to one dimension, the algorithm outlined above can be directly
adapted to the three-dimensional system.  The algorithmic steps of our
event-driven simulation are now:
\begin{enumerate}
\item Start with all the searchers at a distance $r_0$ from the target.
\item For each searcher, with the $i^{\rm th}$ searcher at a radial distance
  $r_i$, draw a random time value from the three-dimensional first-passage
  distribution to a sphere of radius $a$~\cite{R01}:
\begin{subequations}
\begin{equation}
\label{F3d}
  F_{\rm 3d}(r_i,t)=\frac{a}{r_i}\frac{r_i-a}{\sqrt{4\pi D t^3}}\,\, e^{-(r_i-a)^2/4Dt}\,.
\end{equation}
Also choose a reset time $t_r$ from a Poisson distribution with reset rate
$Nr$.

\item If the minimum among these $N+1$ times is the reset time, choose a
  random searcher and reset it to $r_0$.  Each of the remaining searchers is
  moved from its current position $r_i$ to a new position that is drawn from
  the conditional probability 
\begin{equation}
\label{P3d}
     P_{\rm 3d}(r_i,t_r) = \frac{r_i}{r_0}\frac{1}{\sqrt{4 \pi D t_r}} 
\left[e^{-(r_i-r_0)^2/4 D t_r} - e^{-(r_i+r_0-2a)^2/4 D t_r}\right]
\Big/S(r_0,t_r)\,,
\end{equation}
with survival probability now equal to~\cite{R01}
\begin{equation}
\label{S3d}
S_{\rm 3d}(r_0,t_r)= 1-\frac{a}{r_0}\erfc{\bigg(\frac{r_0-a}{\sqrt{4 D t_r}}\bigg)}\,.
\end{equation}
\end{subequations}
Here $P_{\rm 3d}(r_i,t_r)$ corresponds to diffusive propagation of each
searcher over a time $t_r$, subject to the constraint that each searcher
cannot reach a spherical target of radius $a$.  After all the searchers are
moved, increment the elapsed time by $t_r$ and return to step (i).

\item If the minimum among the $N+1$ random times is one of the first-passage
  times, then the target is found.  The total search time is the current
  elapsed time plus this first-passage time.
\end{enumerate}

\subsection{Two dimensions}

\begin{figure}[ht]
\centerline{\includegraphics[width=0.4\textwidth]{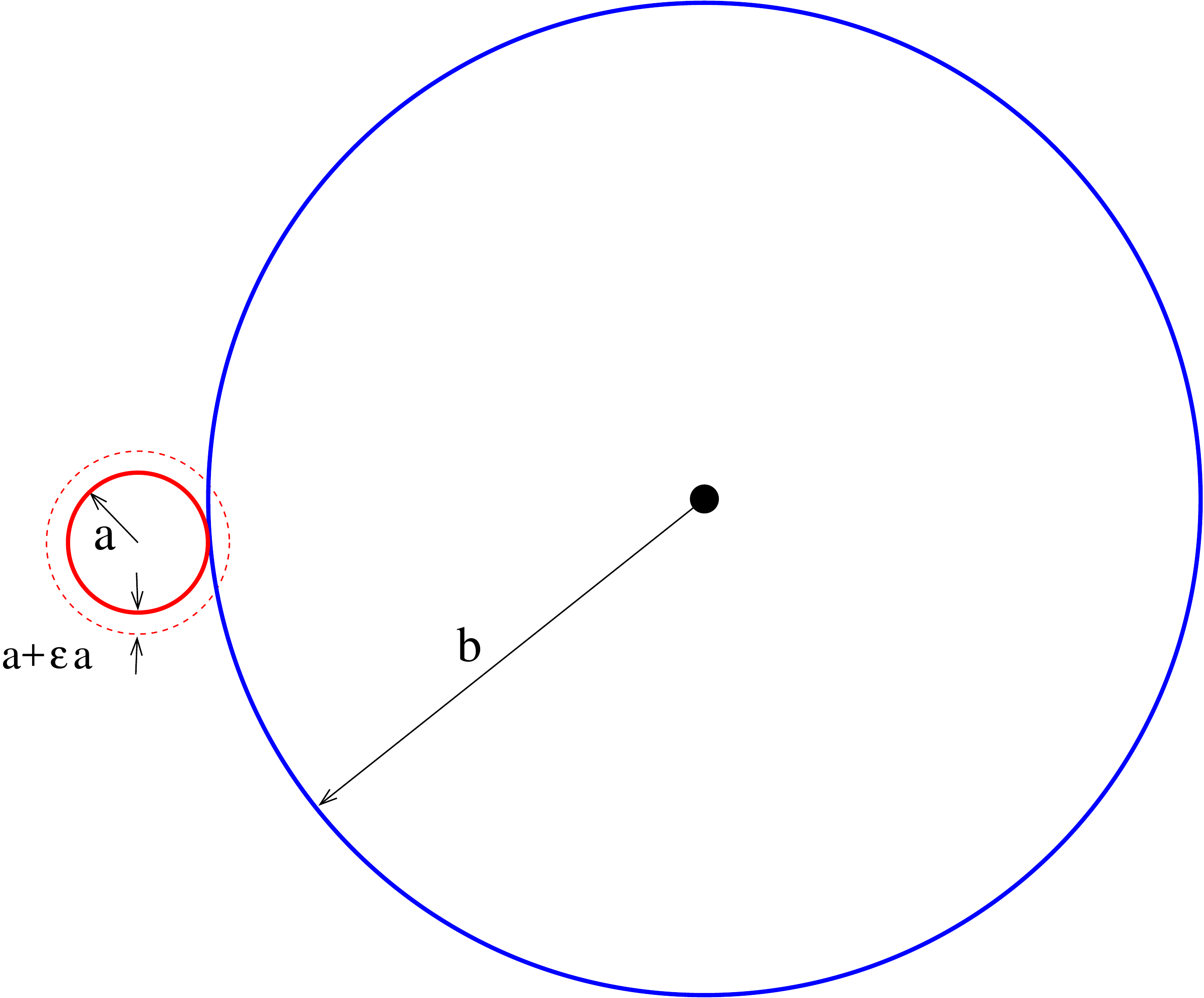}}
\caption{Illustration of a simulation event in two dimensions for a searcher
  that is a distance $b$ from the circumference of a target of radius $a$.}
\label{2d-illust}
\end{figure}

As mentioned in the previously, it is impractical to implement an
event-driven simulation in two dimensions because the exact expression for
the first-passage probability to a circular target of radius $a$ as a
function of time is not known in closed form.  While this first-passage
probability can be expressed as an inverse Laplace transform, the slow
convergence properties of this integral render it not useful as the kernel
for an event-driven simulation.  However, we do know the first-passage
probability in the form of a well-converged series to the circumference of a
circle centered around the current position of the target.  Thus our
simulation is based on propagating the searcher to the circumference of a
circle whose radius adaptively varies depending on the distance to the target
(Fig.~\ref{2d-illust}).  This circle should just touch the target, so that
the radius of this circle is large when the searcher is far from the target
and small when the searcher is close to the target.

For a single searcher that is a distance $b$ from the circumference of the
target, the steps in our algorithm are the following (Fig.~\ref{2d-illust}):

\begin{enumerate}

\item Draw two random times.  One is from the distribution of first-passage times
  $\mathcal{F}(b,t)$
\begin{equation}
  \mathcal{F}(b,t) = \sum_{n=1}^\infty \frac{2}{\mu_n
    {J}_1\left(\mu_n\right)} e^{-\mu_n^2 D t/b^2}
\end{equation}
to the circumference of a circle of radius $b$ (\ref{app:F}).  Here $J_1$ is
the ordinary Bessel function of index 1 and $\mu_n$ is the $n^{\rm th}$ zero
of this Bessel function. Because the jump distance is large if the searcher
is far from the target, little time is spent in simulating the motion of the
searcher when it is wandering aimlessly far from the target.  The second
random time is drawn from the reset time distribution.

\item If the minimum of these two times is the reset time, reset the searcher
  to $r_0$.

\item Otherwise, move the searcher to a random point on the circumference of
  the circle of radius $b$.  If the searcher is within a radius
  $a(1+\epsilon)$ of the center of the target, then we define the target as
  being found.  If the target is not found after the searcher has been moved,
  return to step (i).
\end{enumerate}

We need to introduce an absorbing shell of thickness $\epsilon a$ around the
target to ensure that the searcher actually finds the target.  Clearly, the
apparent search time decreases as $\epsilon$ is increased.  To determine the
appropriate choice of $\epsilon$, we simulate the search with successively
smaller values of $\epsilon$ until the results do not change within the
statistical errors of the simulation and then use the largest of this set of
$\epsilon$ values for simulational efficiency.  For the case of
$a/r_0=10^{-1}$, this value is $\epsilon=10^{-3}$.

By averaging over many trajectories, we construct an accurate numerical
estimate for survival probability due to a single searcher, $S_1(t)$.  Due to
the independence of the searchers, we construct the survival probability for
$N$ searchers by $S_N(t) = \big[S_1(t)\big]^N$.  The average search time is
then given by $\left\langle t_N \right\rangle = \int_0^\infty S_N(t)dt$.

\section{Deterministic Resetting}
\label{sec:DR}

We now investigate the situation where all searchers are reset to their
starting point after a fixed time $T$.  As we shall see, this deterministic
resetting typically leads to a more efficient search compared to stochastic
reset.  Moreover, because all searchers are reset simultaneously, we are able
to obtain numerically exact results for the search time for deterministic
resetting in both one and three dimensions.

\subsection{One dimension}

\subsubsection{Single searcher.}\,\, We follow the renewal approach of
Sec.~\ref{sec:SR1d} to calculate the average search time for a single
searcher that is reset to $x_0$ after a fixed time $T$.  In this case $R(t)$,
the probability that reset time is greater than $t$, is just the Heaviside
step function $H(T-t)$, where $H(z)=1$ for $z>0$ and $H(z)=0$ for $z<0$.
From Eqs.~\eqref{q} and \eqref{TdTr}, we have
\begin{equation}
\begin{split}
\label{QT}
Q=-\int_0^\infty dt\, S(x_0,t)\, \frac{dR(t)}{dt} = S(x_0,T)\,,\\
T_d +T_r = \int_0^\infty dt\, R(t)\, S(x_0,t)= \int_0^T S(x_0,t)\, dt \,.
\end{split}
\end{equation}
Substituting these expressions into Eq.~\eqref{t-recur}, the average search
time for deterministic resetting of a single searcher is
(Fig.~\ref{1D-determ})
\begin{align}
\label{t1-det}
\langle t_1\rangle &= \int_0^T S(x_0,t)\, dt/\big(1-S(x_0,T)\big)\nonumber \\
&=\left[\!\sqrt{\frac{x_0^2\,T }{\pi D}}\, e^{-{x_0^2}/{4DT}}-\frac{x_0^2}{2D}\, 
\text{erfc}\Big(\frac{x_0}{
   \sqrt{4DT}}\Big)+ T \,\mathrm{erf}\Big(\frac{x_0}{\sqrt{4 DT}}\Big)\!\right]\!\bigg/\!
\mathrm{erfc}\Big(\frac{x_0}{\sqrt{4 DT}}\Big)\,.
\end{align}

\subsubsection{Multiple searchers.} 

For multiple searchers, we merely need to use $[S(x_0,t)\big]^N$ rather than
$S(x_0,t)$ in Eq.~\eqref{QT} to account for $N$ independent searchers that
all must have their hitting time exceed a given threshold.  Thus we have
\begin{equation}
\begin{split}
\label{QTN}
Q=-\int_0^\infty dt\, \big[S(x_0,t)]^N\, \frac{dR(t)}{dt} = \big[S(x_0,T)\big]^N\,,\\
T_d +T_r = \int_0^\infty dt\, R(t)\, \big[S(x_0,t)\big]N= \int_0^T \big[S(x_0,t)\big]^N\, dt \,.
\end{split}
\end{equation}
Substituting these in \eqref{t-recur}, the average search time for $N$
searchers with deterministic reset in one dimension has the simple form
\begin{equation}
\label{tN-det}
\left\langle t_N \right\rangle = \frac{\int_0^T \big[S(x_0,t)\big]^Ndt}
{1 - \big[S(x_0,T)\big]^N }\,.
\end{equation}
While we can compute the integral in \eqref{tN-det} analytically for the
cases $N=1,2$ and $N\to\infty$ (\ref{t2inf}), Mathematica can perform the
integration numerically to arbitrary precision to give $\langle t_N\rangle$
for any $N$.

\begin{figure}[ht]
\centerline{\subfigure[]{\includegraphics[width=0.48\textwidth]{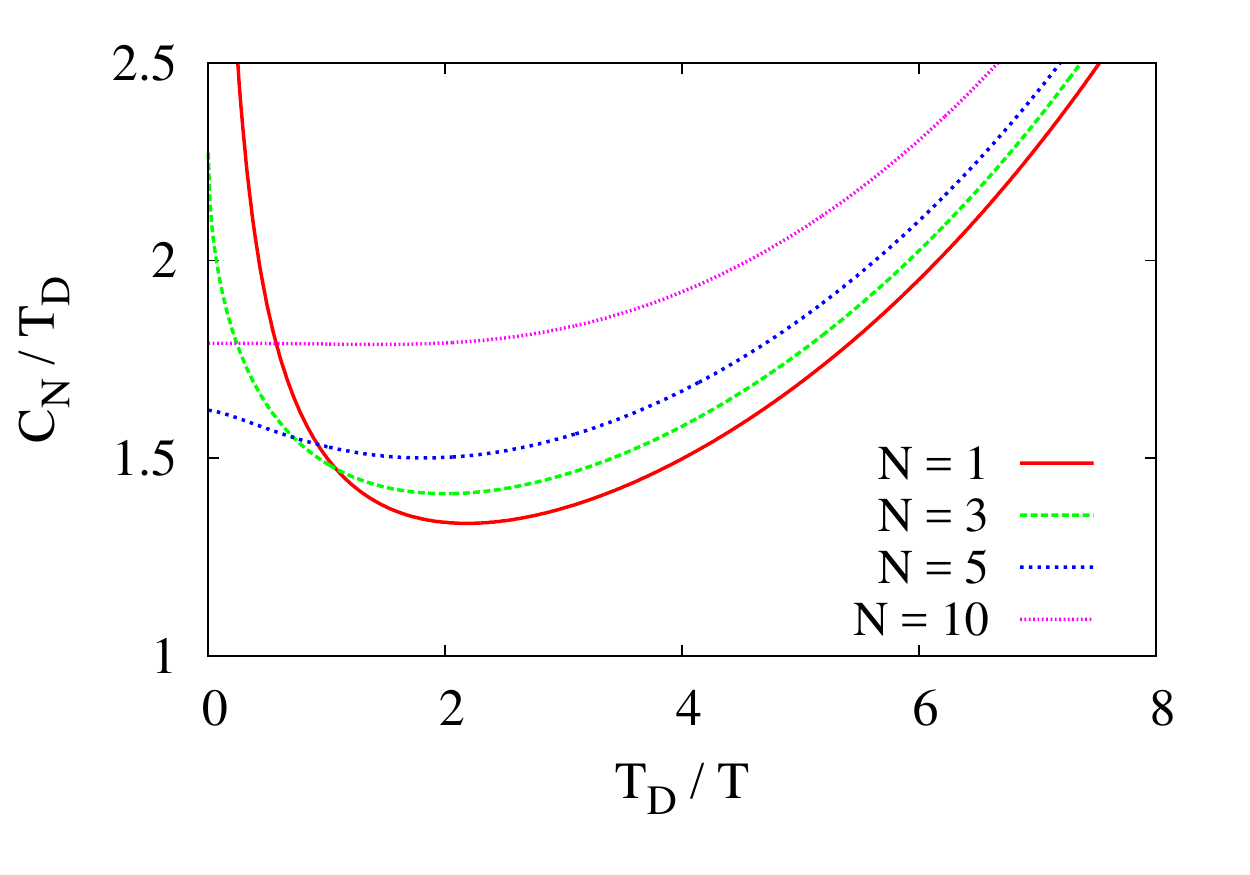}}
\subfigure[]{\includegraphics[width=0.48\textwidth]{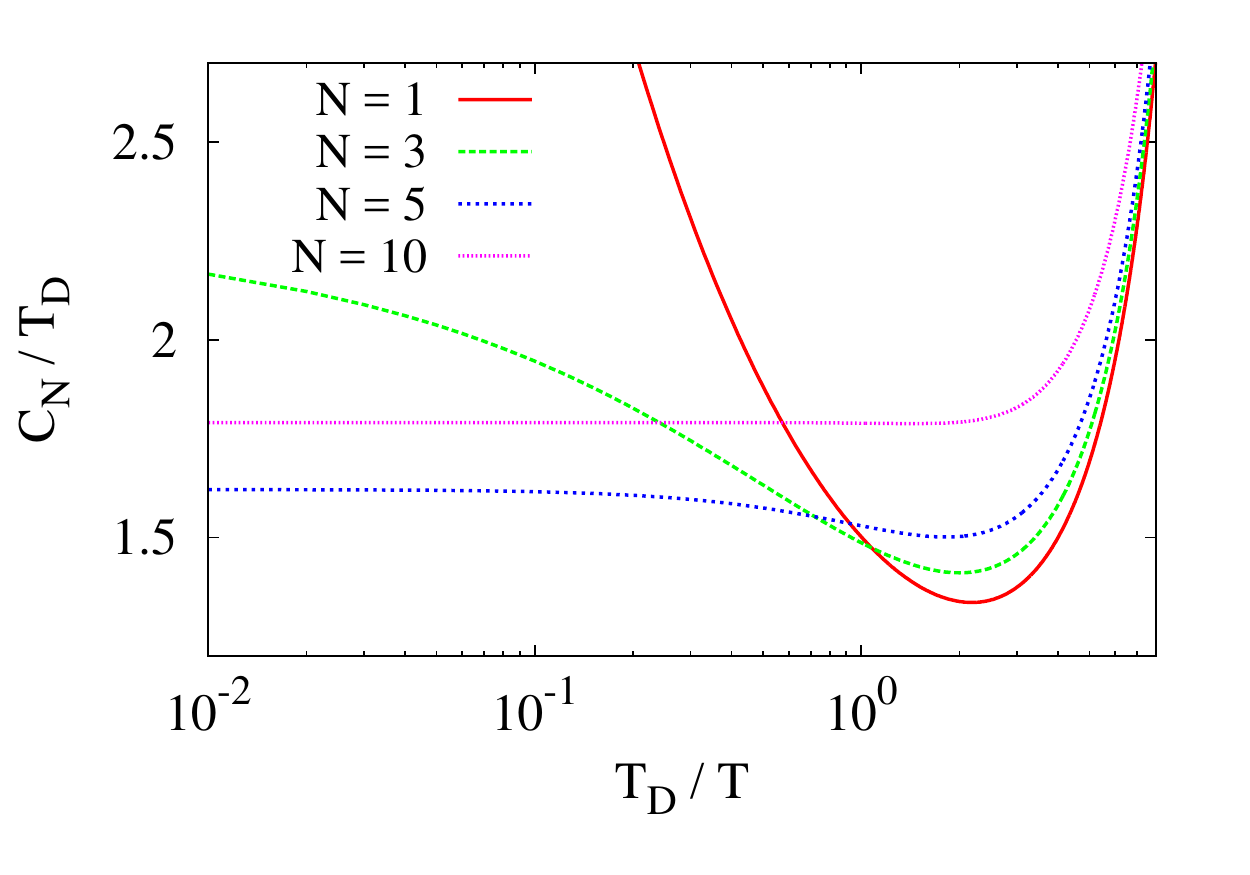}}}
\caption{Average search cost scaled by the diffusion time $T_D$ versus
  $T_D/T$ in one dimension for various $N$ and for deterministic resetting.
  The abscissa scale is linear in (a) and logarithmic in (b).}
\label{1D-determ}
\end{figure}

Figure~\ref{1D-determ} shows the search cost as a function of $1/T$ for
representative $N$ values.  We plot the search cost versus $1/T$ because
$1/T$ plays the same role as the rate $r$ in stochastic resetting.  Several
new features of the search cost for deterministic reset in one dimension
are worth emphasizing:
\begin{enumerate}

\item The lowest search cost is achieved by a single searcher (as in
  stochastic reset) in which the optimal scaled reset time is
  $T_1^*/T_D\approx 0.458$, leading to an optimal scaled search time
  $\langle t_1\rangle \approx 1.336\, T_D$, compared to the optimal cost
  $1.544\, T_D$ from stochastic resetting.  This optimal time corresponds to
  approximately 3 reset events before the target is found.

\item For large $N$, the search cost becomes nearly independent of $1/T$ over
  a wide range (Fig.~\ref{1D-determ}(b)).  This behavior is characterized by
  progressively more derivatives of the cost with respect to $1/T$ becoming
  zero as $1/T\to 0$. In particular, the first derivative is zero for $N>4$,
  the second is zero for $N>6$ and the third is zero for $N>8$.  In general,
  we find that the $k^{\rm th}$ derivative becomes zero for $N= 2k+3$.  We
  can understand this pattern of behavior by differentiating
  Eq.~\eqref{tN-det} with respect to $1/T$, using Mathematica, to find the
  leading behavior
  \begin{align}
    \frac{\partial \left\langle t_N \right\rangle}{\partial (1/T)} 
    &\simeq A_1 T^{3/2}\Big( \erf{\sqrt{{x_0^2 }/{4 D T}}} \Big)^{N-1} 
      - A_2 T^2 \Big( \erf{\sqrt{{x_0^2 }/{4 D T}}} \Big)^{N}
  \end{align}
  where $A_1$ and $A_2$ are $\mathcal{O}(1)$ in $T$.  As $1/T\rightarrow 0$,
  the error function is proportional to its argument, so that the above
  leading terms scale as a negative power of $T$ for $N>4$ and as a positive
  power for $N<4$.  Analogous behavior arises for higher derivatives.
  
\begin{figure}[ht]
\centerline{\includegraphics[width=0.55\textwidth]{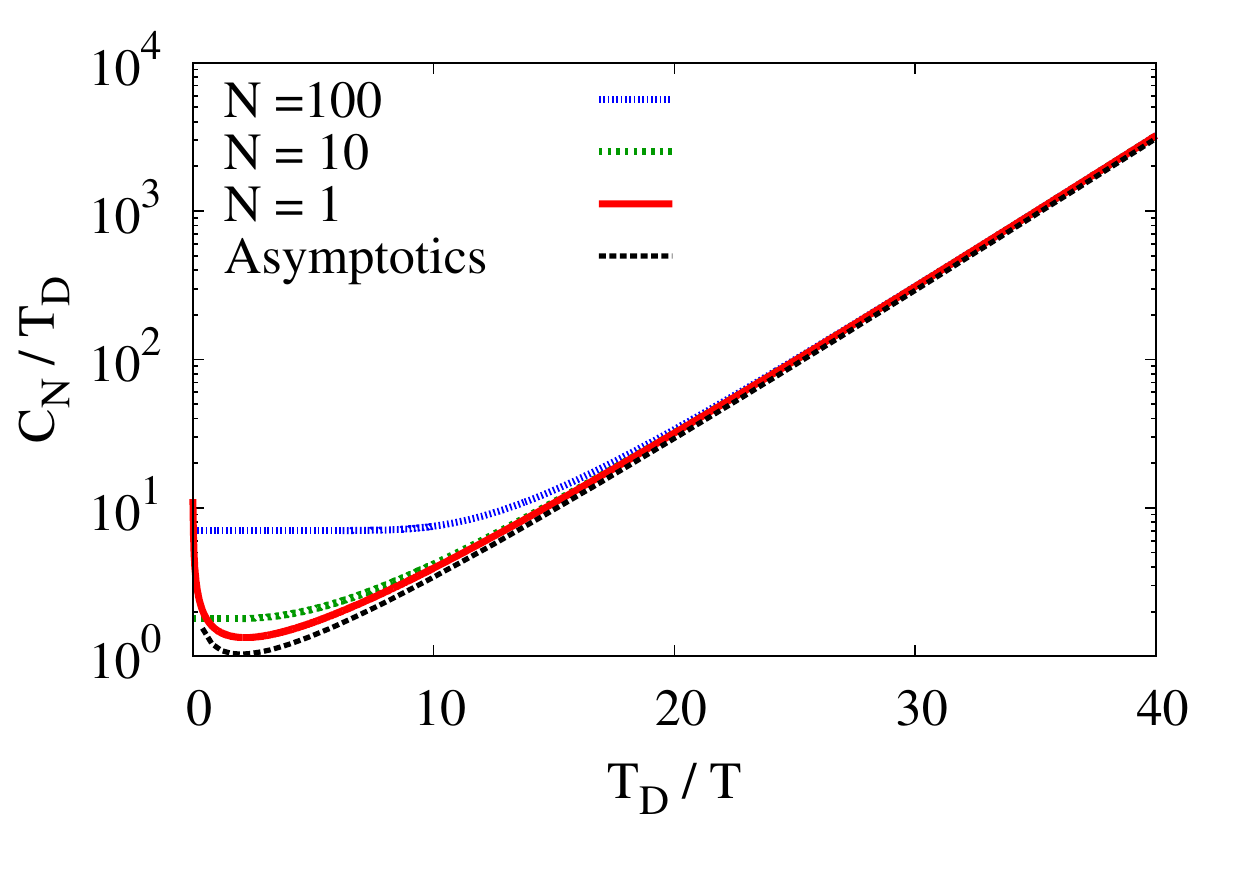}}
\caption{Comparison between the exact formula \eqref{tN-det} and the
    asymptotic formula \eqref{tN-asymp} for the mean hitting time in one
    dimension with deterministic reset.}
\label{1D-determ-compare}
\end{figure}

\item For $T_D/T\to\infty$, the search time asymptotically increases as
  $e^{T_D/T}$, whereas for stochastic resetting the corresponding
  $r\to\infty$ behavior is the search time growing as $e^{\sqrt{r}}$.  The
  $e^{T_D/T}$ growth has a simple origin.  As $T_D/T\to\infty$, the probability
  that one searcher does not reach the target within the reset time $T$ is
  $S=\text{erf}(x_0/\sqrt{4DT})$.  The probability that none of the searchers
  reaches the target within time $T$ is $S^N$, so that the probability that
  at least one of the searchers reaches with time $T$ is $1-S^N$.  In the
  limit $T_D/T\to\infty$, the asymptotic behavior of this probability is
\begin{align*}
1-S^N\simeq \frac{\sqrt{4DT}\, N\, e^{-x^2/4DT}}{\sqrt{\pi}\, x_0}\ll 1\,.
\end{align*}
The number of reset events until a searcher finds the target is the inverse
of this expression.  Multiplying by $NT$ gives the asymptotic behavior of the
average search cost
\begin{equation}
\label{tN-asymp}
  \langle C_N\rangle \simeq \frac{NT}{1-S^N}\simeq \,T\,\sqrt{\frac{\pi x_0^2}{4DT}} \,\,e^{x_0^2/4DT}\,.
\end{equation}
This asymptotics matches the numerically exact expression for the search time
when $T_D/T\gg 1$ (Fig.~\ref{1D-determ-compare})

\end{enumerate}

\subsection{Three dimensions}
\label{det-3d}

In three dimensions, we again obtain numerically exact results for
deterministic reset for all parameter values and thus can probe the role of
the reset time, as well as the parameter $a/r_0$ on the search cost and
search time.  While the qualitative dependence of the search cost and time on
$1/T$ is the same for all $a/r_0$, there are quantitative anomalies that are
worth highlighting.

\begin{figure}[ht]
\centerline{\includegraphics[width=1.\textwidth]{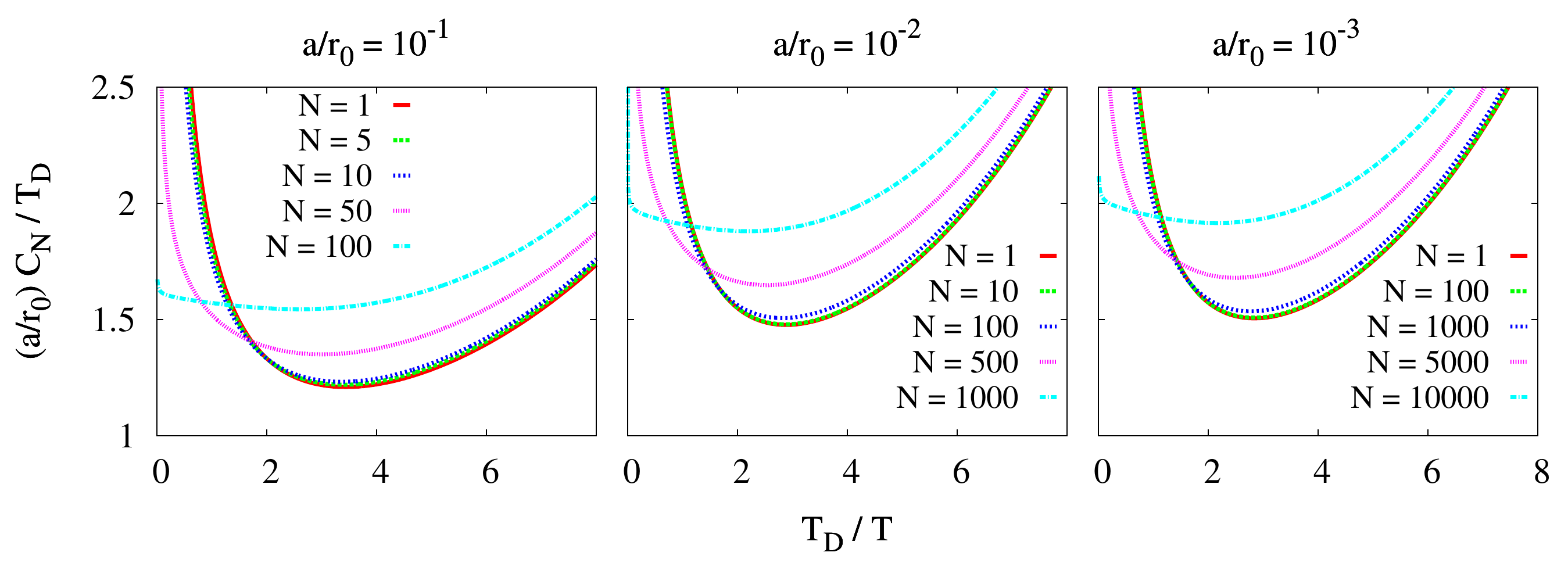}}
\caption{Average search cost scaled by $aT_D/r_0$ versus $T_D/T$ in three
  dimensions for various $N$ and $a/r_0$. }
\label{3D-determ}
\end{figure}

We again use the correspondence between one-dimensional and three-dimensional
diffusion to determine the search time in three dimensions.  In the renewal
formula~\eqref{tN-det} for the search time, we now need the first-passage and
survival probabilities in three dimensions, $F_{\mathrm{3d}}(r_0,t)$ and
$ S_{\mathrm{3d}}(r_0,t)$, respectively (Eqs.~\eqref{F3d} and \eqref{S3d}).
Substituting these expressions into Eq.~\eqref{tN-det} and using Mathematica
to perform the integrals numerically, we again obtain the search time and
cost as a function of $1/T$ for any $N$ with arbitrary precision
(Fig.~\ref{3D-determ}).

As expected, the search cost is initially a decreasing function of $T_D/T$ for
any $N$.  When $T_D/T=0$, a diffusing searcher is transient in three dimensions
and does not necessarily find the target; thus the search cost diverges in
this limiting case, even when $N$ is large.  On the other hand, for
$T_D/T\to\infty$, the search again becomes inefficient because each searcher is
typically reset before it can progress towards the target.

Figure~\ref{3D-determ} also illustrates a data collapse within each panel
when $N \ll r_0/a$ and between panels in the limit of $r_0/a\gg1$.  This
implies that the search cost is independent of $N$ and $a/r_0$, for small
enough target size and number of searchers, when this cost is scaled by
$a /(r_0 T_D)$.  To derive this behavior, we substitute Eqs.~\eqref{F3d} and
\eqref{S3d} into \eqref{tN-det}, and approximate the argument of the error
function, $(r_0-a)/\sqrt{4 D T}$, as $\sqrt{T_D / 4 T}$.  These steps lead to
\begin{equation*}
\left\langle t_N \right\rangle = \frac{\int_0^T \left[1-(a/r_0)\erfc{\Big(\sqrt{T_D/4 T}\Big)}\right]^Ndt}{1-\left[1-(a/r_0)\erfc{\Big(\sqrt{T_D/4 T}\Big)}\right]^N}~.
\end{equation*}
In the limit $r_0 / (a N) \gg 1$, the search cost
$\langle C_N\rangle = N\langle t_N\rangle$ is
\begin{equation}
\label{tN-det-approx-1}
\left\langle C_N \right\rangle = \frac{r_0 T}{a}\left[\erfc{\left(\sqrt{{T_D}/{4 T}}\right)}\right]^{-1}\,.
\end{equation}
This scaling form implies that plots of $\langle C_N\rangle$ versus $1/T$
collapse onto a single curve when $a \left\langle C_N \right\rangle/ r_0$ is
plotted against $T_D/T$.  The asymptotic form~\eqref{tN-det-approx-1} may
also be derived by merely counting the number of resets until the target is
found.  The probability of hitting the target within a single reset event by
a single searcher is given by,
\begin{equation*}
p \equiv 1 - S(r_0,T) = \frac{a}{r_0}\erfc{\left(\frac{r_0-a}{\sqrt{4 D
        T}}\right)}\,.
\end{equation*}
The probability that any of the $N$ searchers finds the target is
$1\!-\!(1\!-\!p)^N\simeq N p \equiv P$. Since each of these hitting events is
independent, the average number of reset events before one of the searchers
reaches the target is
$\sum_{n\geq 1} n P (1-P)^n = 1/P\simeq r_0 / \big[N a \erfc(\sqrt{{T_D}/{4
    T}})\big]$.
In the limit of large number of resets, the search time is just $T$ times
number of resets, as the time for the last segment of the trajectory that
actually reaches the target is negligible.  This reasoning again leads to
Eq.~\eqref{tN-det-approx-1}.

\subsection{Two dimensions}

\begin{figure}[ht]
\centerline{\subfigure[]{\includegraphics[width=0.50\textwidth]{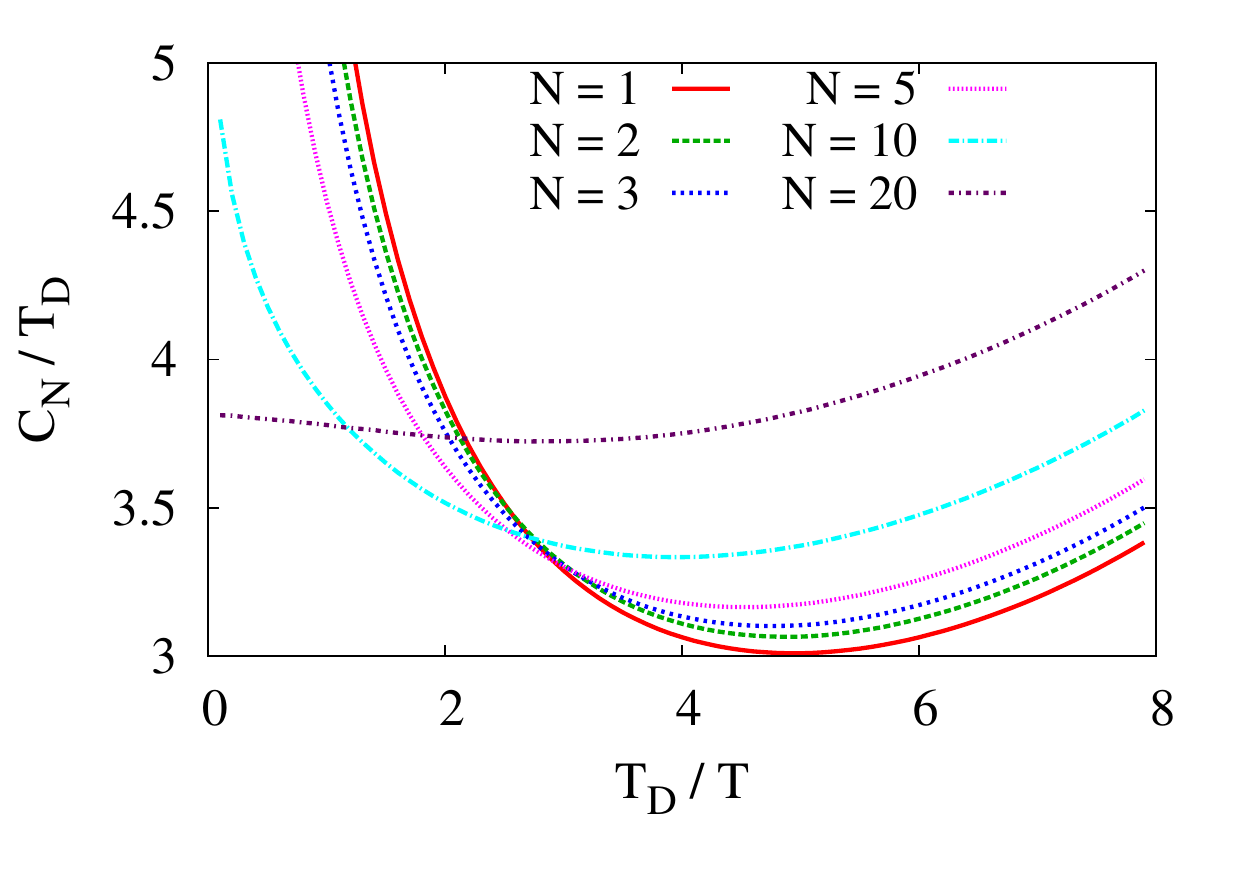}}
\subfigure[]{\includegraphics[width=0.48\textwidth]{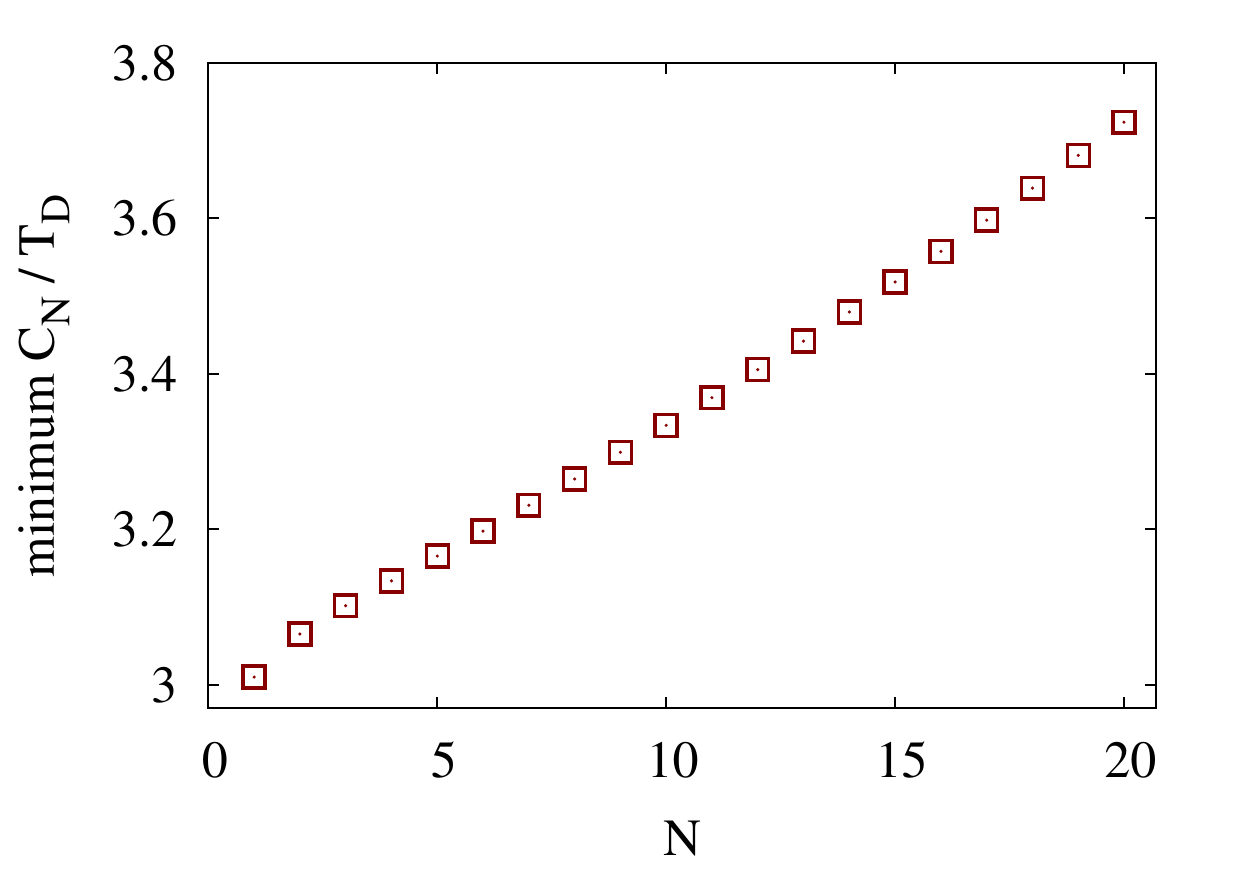}}}
\caption{(a) Average scaled search cost $C_N/T_D$ versus scaled inverse reset
  time $T_D/T$ in two dimensions for various $N$ and for $a/r_0=10^{-1}$.
  Data are averaged over $10^8$ trajectories.  (b) Minimum search cost as a
  function of $N$. }
\label{2d-determ}
\end{figure}

In two dimensions however, we are not able to implement the renewal process
calculation, as we do not have exact expressions for $S(r_0, t)$ or
$F(r_0, t)$. Hence, we use the same simulation as in the stochastic reset in
$d=2$.  As in one and three dimensions, minimum cost is achieved for $N=1$
searcher and the cost monotonically increases with $N$
(Fig.~\ref{2d-determ}).

\section{Discussion}
\label{sec:disc}

In this work, we explored the consequences of superimposed resetting on the
performance of stochastic search processes.  In this resetting, either one or
many searchers are returned to a fixed home base either at a fixed rate or at
a fixed reset time.  We found a variety of intriguing and sometimes
unexpected features.  When each searcher is independently reset to the home
base at a fixed rate $r$, the search cost is minimal for a single searcher
when the reset rate is of the order of the inverse diffusion time
$T_D=x_0^2/D$.  We also found that resetting always hinders the search for
$N\geq 8$ searchers, while for $N\leq 7$, there is an optimal nonzero reset
rate for each $N$.

By exploiting the well-known relation between the diffusion equation in one
and in three dimensions, we obtained analogous results for search with
stochastic resetting in three dimensions.  The primary new result for $d=3$
is that the cost of the search is nearly independent of the number of
searchers for $N<10$ for the case of $a/r_0 = 10^{-1}$.  In two dimensions,
we developed an alternative procedure in which we simulate the target
survival probability in the presence of a single searcher and then take the
$N^{\rm th}$ power of this quantity to obtain the target survival probability
in the presence of $N$ searchers.  In this case, similar to three dimensions,
the cost is nearly independent of the number of searchers up to $N=10$ and
for $N>10$, the cost increases monotonically.

We also explored a related model in which all searchers are simultaneously
reset to the home base after a fixed operation time $T$.  This deterministic
resetting is theoretically and computationally simpler than stochastic
resetting, and we are able to obtain explicit formulae for the average search
time for any $N$ that can be numerically integrated to arbitrary precision.
In one dimension, deterministic resetting gives a search time that becomes
independent of the reset time $T$ when $N$ is sufficiently large, a behavior
that can be understood from the small-$r$ behavior of the search time.  In
three dimensions, we showed by a simple extremal argument that the search
cost versus $1/T$ becomes independent of $N$.

There are a wide range of extensions of the basic model to practically and
theoretically interesting situations.  It would be worthwhile to extend
search with resetting to the cases where either the target is diffusing
and/or the target is mortal~\cite{YAL13,AYL13,CAMYL15,MR15}.  These
generalizations would naturally describe, e.g., the occupants of a lifeboat
that is adrift in the ocean.  When the target is also moving, the basic
question is again whether the reset helps or hinders the search. For a mortal
target with any reasonable distribution of mortality, there will always be a
non-zero probability that the target will die before being found and the
relevant issue is to construct appropriate criteria that lead to a
well-defined optimization problem.

Financial support for this research was also provided in part by the grants
DMR-1623243 from the National Science Foundation (UB and SR), from the John
Templeton Foundation (CDB and SR), and Grant No.\ 2012145 from the United
States—Israel Binational Science Foundation (UB).  We also thank B. Meerson
for helpful discussions and S. Reuveni for useful comments on the manuscript.

\emph{Added Note:} As final revisions were being made, we became aware of
related work~\cite{PKE16}, in which the authors mathematically showed that
deterministic reset leads to the smallest search time for the case of a
single searcher, as we also observed.

\appendix

\section{The Probability Distribution}
\label{PD}

The full description for a static target and one searcher that is
stochastically reset to $x_0$ can be obtained from the time-dependent
probability distribution $\rho(x,t)$.  Its evolution is governed by the
diffusion equation, supplemented by terms that account for the resetting:
\begin{equation}
\label{DD}
\rho_t = D\rho_{xx} - r\rho + r\delta(x-x_0)\int_0^\infty \rho(x,t)\,
dx\,.
\end{equation}
Here $r\rho$ accounts for the loss of probability at rate $r$ at position $x$
due to the resetting, while the integral accounts for the gain of probability
at the reset point $x_0$.  The amplitude of this gain term equals the total
probability that the target has not yet been found, which is less than 1 and
also decreasing with time.  We solve this equation, subject to the absorbing
boundary condition $\rho(0,t>0)=0$, corresponding to the loss of probability
whenever the target at the origin is reached.  For simplicity, we consider
the initial condition $\rho(x,t\!=\!0)=\delta(x\!-\!x_0)$.

To solve Eq.~\eqref{DD}, we first Laplace transform it to give
\begin{equation}
\label{DDL}
s\rho -\delta(x-x_0)=D\rho''-r\rho+r\delta(x-x_0)\int_0^\infty\rho(x,s)\,
dx\,.
\end{equation}
Here $\rho=\rho(x,s)$ is the Laplace transform of $\rho(x,t)$ and the prime
denotes differentiation with respect to $x$.  For $x\ne x_0$, we must solve
$\rho''=\left(\frac{s+r}{D}\right)\rho\equiv \alpha^2\rho$, with general
solution
\begin{equation*}
\rho(x,s)= 
\begin{cases}
A\, e^{\alpha x} + B\, e^{-\alpha x} & \qquad x<x_0\,,\\
C\, e^{-\alpha x} & \qquad x>x_0\,.
\end{cases}
\end{equation*}
For $x>x_0$, only the decaying exponential appears so that the probability
distribution does not diverge as $x\to \infty$.

The absorbing boundary condition at the origin immediately gives $A+B=0$,
which simplifies the density in the range $x<x_0$ to
$\rho=A\,\textrm{sinh}\, \alpha x$.  Continuity of the probability
distribution at $x_0$ gives the condition
$A\,\textrm{sinh}\, \alpha x_0=C\, e^{-\alpha x_0}$, which we use to
eliminate $C$.  After some standard and simple steps, the form of $\rho(x,s)$
for $x<x_0$ and $x>x_0$ can be expressed more symmetrically as
\begin{equation}
\label{rho-gen}
\rho(x,s)= 
\begin{cases}
A\, {\displaystyle \frac{\textrm{sinh}\, \alpha x}{\textrm{sinh}\, \alpha x_0}} &
\qquad x<x_0\,,\\[0.2in]
A\, e^{-\alpha (x-x_0)} & \qquad x>x_0\,,
\end{cases}
\end{equation}
which is manifestly continuous at $x=x_0$.

The constant $A$ is determined by the joining condition, which is obtained by
integrating \eqref{DDL} over an infinitesimal region that includes $x_0$:
\begin{equation}
\label{join}
-1 = D\left(\rho_+'-\rho_-'\right)+r \int_0^\infty \rho(x',s)\, dx'\,.
\end{equation}
Here $\rho_+'$ is the gradient of $\rho$ as $x\to x_0^+$ and similarly for
$\rho_-'$.  Using Eq.~\eqref{rho-gen}, we have
\begin{align}
\begin{split}
\rho_+' &= - A\,\alpha\,\, e^{-\alpha(x-x_0)}\Big|_{x=x_0} = -A\,\alpha\,, \\
\rho_-' &= A\,\alpha \,\frac{\textrm{cosh}\, \alpha x}{\textrm{sinh}\, \alpha x_0}\Big|_{x=x_0}= A\,\alpha\,\,
\textrm{coth}\, \alpha x_0\,.
\end{split}
\end{align}
We also need 
\begin{align}
\begin{split}
\label{rho-int}
 \int_0^\infty \rho(x',s)\, dx'&=\int_0^{x_0} A \, \frac{\textrm{sinh}\, \alpha
                                 x'}{\textrm{sinh}\, \alpha x_0}\,dx'
+\int_{x_0}^\infty A\,\, e^{-\alpha (x'-x_0)}\,dx'\,,\\
&= \frac{A}{\alpha}\left[\frac{(\textrm{cosh}\, \alpha x_0-1)}{\textrm{sinh}\,
  \alpha x_0}+1\right]\,.
\end{split}
\end{align}
Substituting the above into the joining condition gives, after
straightforward algebra,
\begin{equation}
\label{A}
A= \frac{\textrm{sinh}\, \alpha x_0}{D\alpha\, e^{\alpha x_0}+\frac{r}{\alpha}\big(1-e^{\alpha
      x_0}\big)}\,.
\end{equation}
This, together with Eq.~\eqref{rho-gen}, gives the probability distribution
in the Laplace domain.

From this solution, the Laplace transform of the flux to the origin is:
\begin{align}
\label{j0s}
j(0,s) &= D\rho(x,s)'\Big|_{x=0}= \frac{DA\alpha}{\textrm{sinh}\, \alpha
  x_0}\,,\nonumber \\[0.1in]
&= \left[ e^{\alpha x_0} +\frac{r}{D\alpha^2}\big(1-e^{\alpha x_0}\big)\right]^{-1}\,.
\end{align}
By definition, $j(0,s)$ is also the moment generating function
\begin{align}
j(0,s)&=\int_0^\infty j(0,t)\, e^{-st}\, dt\,,\nonumber \\
&= \int_0^\infty j(0,t)\big[1-st +\tfrac{1}{2} (st)^2-\ldots\big]\,,\nonumber \\[0.1in]
&= 1 - s\langle t\rangle +\tfrac{1}{2}s^2 \langle t^2\rangle -\ldots\,.
\end{align}
Expanding \eqref{j0s} in a power series in $s$, we obtain the results quoted
in Eq.~\eqref{t}.

As a byproduct, the survival probability of the searcher in the Laplace
domain, $S_1(x_0,s)$ (which coincides with the survival probability of the
target) is the spatial integral of $\rho(x,s)$ in Eq.~\eqref{rho-int}.  Using
\eqref{A}, and after some simple algebra, we obtain
 \begin{equation}
\label{S1s}
S_1(x_0,s) = \frac{1 - e^{-\alpha x_0}}{s + r e^{-\alpha x_0}}\,,
\end{equation}
as first obtained in Ref.~\cite{EM11a,EM11b} by different means.

\section{Slope of $\langle t_N\rangle$ for $r\to 0$}
\label{7-8}

Because of the independence of the searchers, the probability that the target
is not found by $N$ searchers within time $t$, $S_N(t)$, is just
$\big[S_1(t)\big]^N$.  Thus the slope of the average search time at $r=0$ is
\begin{align}
\label{dtdr}
\frac{d \langle t_N\rangle}{d r}\bigg|_{r=0} 
&=\frac{d}{d r}\int_0^\infty\big[S_1(t)\big]^N\,dt\,\,\bigg|_{r=0}\nonumber \\ 
 &= \int_0^\infty N\,\, \big[S_1(t,r\!=\!0)\big]^{N-1} \,\,
\frac{d S_1(t,r)}{d r}\,\,dt\,\bigg|_{r=0}
\end{align}
Since we just need the slope at $r=0$, we expand the Laplace transform
$S_1(x_0,s)$ in \eqref{S1s} to first order in $r$ to give
\begin{equation}
\label{S1ssmallr}
\lim_{r\to 0} S_1(x_0,s) = \frac{1}{s}\left[1-e^{-x_0 \sqrt{{s}/{D}}} + \frac{r}{s} 
\,e^{-x_0 \sqrt{{s}/{D}}}\left(x_0 \sqrt{{s}/{4 D}} - 1 + e^{-x_0 \sqrt{{s}/{4D}}}\right)\right]
\end{equation}
Using Mathematica, the Laplace inverse of the above expression is
\begin{align}
\label{S1tsmallr}
\lim_{r\to 0}  S_1(x_0,t) =  &\frac{r x_0^2}{D} + 2  r x_0 \sqrt{\frac{t}{\pi D}}
                     \left(e^{-{x_0^2}/{4 D t}} - e^{-{x_0^2}/{D t}}\right) \nonumber \\ 
     &+ \left(1 +  rt + \frac{r x_0^2}{D}\right)\mathrm{erf}\left(\frac{x_0}{\sqrt{4Dt}}\right)
  - \left(r t + \frac{2 r x_0^2}{D}\right)\mathrm{erf}\left(\frac{x_0}{\sqrt{D t}}\right)\,.
\end{align}
Differentiating Eq.~\eqref{S1tsmallr} with respect to $r$ and substituting
in~\eqref{dtdr} gives
\begin{align}
\frac{d \langle t_N\rangle}{d r}\bigg|_{r=0}\!\!\! = N \,T_D^2 
\int_0^\infty \!\left[\mathrm{erf}{\Big(\frac{1}{\sqrt{4 \tau}}\Big)}
  \right]^{N-1}&
\!\left\{\!1 + \sqrt{\frac{4 \tau}{\pi}}\left(e^{-{1}/{4\tau}}-e^{-{1}/{\tau}}\right) \right. \nonumber \\
 &+ \left. (\tau\!+\!1)\,\mathrm{erf}{\Big(\frac{1}{\sqrt{4\tau}}\Big)} 
- (\tau\!+\!2)\,\mathrm{erf}{\Big(\frac{1}{\sqrt{\tau}}\Big)}\!\right\} d\tau
\end{align}
where again $T_D = x_0^2/D$.  Evaluating the integral numerically shows that
the initial slope changes sign at $N\approx 7.326477\ldots$ (see
Fig.~\ref{fig:dtdr}).  Thus for $N\leq 7$ searchers, resetting at a non-zero
optimal rate speeds up the search compared to no resetting, while for
$N\geq 8$, resetting always hinders the search.

\section{Survival probability from an absorbing circle}
\label{app:F}
To find the survival probability from an absorbing circle of radius $R$ centered around the origin, with the initial condition $r_0=r$, we begin with the diffusion equation in two dimensions,
\begin{equation}
\frac{\partial S(r,t)}{\partial t} = D\nabla^2 S(r,t) = D\left( \frac{\partial^2 S(r,t)}{\partial r^2} + \frac{1}{r}\frac{\partial S(r,t)}{\partial r} \right)
\end{equation}
Due to circular symmetry, the survival probability is independent of the polar angle, and so we have kept only the radial term of the Laplacian operator in 2d. Assuming separation of variables and defining $S(r,t)=\mathcal{R}(r)\mathcal{T}(t)$ and re-arranging the terms, we get
\begin{equation}
\frac{\dot{\mathcal{T}}}{D \mathcal{T}} = \frac{\mathcal{R}''}{\mathcal{R}} + \frac{1}{r}\frac{\mathcal{R}'}{\mathcal{R}}
\end{equation}
where the over-dot refers to derivative w.r.t $t$ and prime refers to derivative w.r.t $r$. Equating both sides to a negative constant $-\mu^2$, we obtain
\begin{equation}
\mathcal{T}(t) = e^{-\mu^2 D t} \quad \text{and} \quad r^2 \mathcal{R}'' + r \mathcal{R}' - \mu^2 r^2 = 0
\end{equation}
The equation for $\mathcal{R}(r)$ is in the form of a Bessel equation of order $0$, giving us the general solution,
\begin{equation}
S(r,t) = \sum_\mu A_\mu J_0(\mu r) e^{-\mu^2 D t}
\end{equation}
Applying the absorbing boundary condition, $S(r=R,t)=0$, we get $\mu = \mu_n /R$ where $\left\lbrace\mu_n\right\rbrace$ are the zeroes of $J_0$
\begin{equation}
\label{Srt}
S(r,t) = \sum_n A_n J_0\left(\frac{\mu_n r}{R}\right) e^{-\mu^2 D t}
\end{equation}
To calculate the coefficients $A_n$, we use the orthogonality condition of
the Bessel functions~\cite{AS72}
and the initial condition $S(r,t=0)=1\,\forall r<R$.  Using Eq.~\eqref{Srt} at $t=0$ and multiplying both sides by $(r/R) J_0(\mu_m r/R)$ and integrating with respect to $r/R$ we get,
\begin{equation}
\int_0^1 \frac{r}{R} J_0\left(\mu_m\frac{r}{R}\right)d\left(\frac{r}{R}\right) = \sum_n \int_0^1  A_n \frac{r}{R} J_0\left(\mu_m\frac{r}{R}\right) J_0\left(\frac{\mu_n r}{R}\right) d\left(\frac{r}{R}\right) = \frac{A_n}{2}\left[ J_1\left(\mu_n\right)\right]^2
\end{equation}
Integrating the left-hand side with Mathematica and re-arranging terms, we get,
\begin{equation}
A_n = \frac{2}{\mu_n J_1(\mu_n)}
\end{equation}
Finally, we require the survival probability when starting from the center of the absorbing circle, so substituting $r=0$, we get
\begin{equation}
S(r=0,t) = \sum_n \frac{2}{\mu_n J_1(\mu_n)} e^{-\mu_n^2 D t / R^2}
\end{equation}

\section{Search Time for Deterministic Reset for $N=2$ and $N=\infty$}
\label{t2inf}

We start with the general expression \eqref{tN-det} for the average search
time in deterministic search:
\begin{equation}
\label{C1}
\left\langle t_N \right\rangle = \frac{\int_0^T \big[S(t)\big]^Ndt}
{1 - \big[S(T)\big]^N }\,.
\end{equation}
In terms of $u\equiv x_0^2/4DT=T_D/4T$, the integral
$\int_0^T \big[S(t)\big]^Ndt$ can be written as
\begin{align}
\label{I}
\frac{x_0^2}{2D}\int_u^\infty \frac{\text{erf}(z)^N}{z^3}\, dz \equiv \tfrac{1}{2}\,T_D\, I\,.
\end{align}
Repeatedly integrating by parts to reduce the power of the factor in the
denominator, we obtain
\begin{align}
I= \Big(1\!+\! \frac{1}{2u^2}\Big)\text{erf}(u)^N \!+\! \frac{N}{\sqrt{\pi}\,
  u}e^{-u^2}\text{erf}(u)^{N-1}
\!-\!1\!+\!\frac{2}{\pi}N(N\!-\!1)\!\int_u^\infty\!\!\!\text{erf}(z)^{N-2}\,
  e^{-2z^2}\,\frac{dz}{z}\,.
\end{align}
For the case of $N=2$, the last integral is
\begin{equation*}
\int_u^\infty e^{-2z^2}\, \frac{dz}{z}=\Gamma(0,2u^2)\,,
\end{equation*}
where $\Gamma(a,b)$ is the incomplete Gamma function~\cite{AS72}.  Assembling
the above results, the average search time for $N=2$ searchers is
\begin{align}
\langle t_2\rangle =\frac{T_D}{2}\,\frac{\displaystyle{\left[\Big(1+\frac{1}{2u^2}\Big)\text{erf}(u)^2 + \frac{2}{\sqrt{\pi}\,
  u}e^{-u^2}\text{erf}(u)
-1+\frac{2}{\pi}\Gamma(0,2u^2)\right]}}{1-\text{erf}(u)^2}\,.
\end{align}

In the limit $N\to\infty$, the search time becomes arbitrarily small so that
eventually the reset time $T$ is larger than the search time.  In this limit,
we may set $T=\infty$, or equivalently, $u=0$ in Eq.~\eqref{I}.  Thus
Eq.~\eqref{C1} reduces to~\cite{MR15}
\begin{align}
\left\langle t_N \right\rangle =\frac{T_D}{2}\int_0^\infty
  \frac{\text{erf}(z)^N}{z^3}\, dz \simeq \frac{T_D}{4\ln N}\,.
\end{align}
In this limiting case, the average search time no longer depends on the reset
time.

\bigskip\bigskip\bigskip\bigskip


\end{document}